\documentclass[a4paper,fleqn]{cas-sc}

\usepackage[numbers, sort&compress]{natbib}
\usepackage{bbm}
\usepackage{bm}
\usepackage{placeins}
\usepackage{subcaption}
\usepackage{circuitikz}

\usepackage{nomencl}
\makenomenclature
\usepackage{etoolbox}
\renewcommand\nomgroup[1]{\vspace{6pt}%
  \item[\bfseries
  \ifstrequal{#1}{M}{Mathematical symbols}{%
  \ifstrequal{#1}{F}{Fiber orientation-related symbols}{}}%
]}

\def\tsc#1{\csdef{#1}{\textsc{\lowercase{#1}}\xspace}}
\tsc{WGM}
\tsc{QE}

\newcommand{\R}{\mathbb{R}}

\newcommand{\N}{\mathbb{N}}
\newcommand{\E}{\mathbb{E}}
\newcommand{\V}{\mathbb{V}}

\newtheorem{lemma}{Lemma}
\newtheorem{remark}{Remark}

\DeclareMathOperator*{\argmin}{arg\,min}

\begin{document}
\let\WriteBookmarks\relax
\def\floatpagepagefraction{1}
\def\textpagefraction{.001}

\shorttitle{}
\shortauthors{S. Salatovic et~al.}

\title [mode = title]{Reliable Uncertainty Quantification for Fiber Orientation in Composite Molding Processes using Multilevel Polynomial Surrogates}

\let\printorcid\relax

\author[1, 2]{Stjepan Salatovic}
\ead{stjepan.salatovic@kit.edu}
\credit{Writing - original draft, Validation, Software, Methodology, Conceptualization}
\cormark[1]

\author[1]{Sebastian Krumscheid}
\credit{Methodology, Conceptualization, Resources, Writing - review and editing}

\author[2]{Florian Wittemann}
\credit{Conceptualization, Resources, Writing - original draft, Writing - review and editing}

\author[2]{Luise Kärger}
\credit{Conceptualization, Resources, Writing - review and editing}

\cortext[cor1]{Corresponding author}

\affiliation[1]{
    organization={Scientific Computing Center -- Uncertainty Quantification, Karlsruhe Institute of Technology (KIT)},
    citysep={},
    country={Germany},
}

\affiliation[2]{
    organization={Institute of Vehicle System Technology -- Lightweight Technology, Karlsruhe Institute of Technology (KIT)},
    citysep={},
    country={Germany},
}

\begin{abstract}
Fiber orientation is decisive for the mechanical performance of composite materials.
During manufacturing, variations in material and process parameters can influence fiber orientation.
We employ multilevel polynomial surrogates to model the propagation of uncertain material properties in the injection molding process.
To ensure reliable uncertainty quantification, a key focus is deriving novel error bounds for statistical measures of a quantity of interest.
Numerical experiments employ the Cross-WLF viscosity model and Hagen-Poiseuille flow to investigate the impact of uncertainties in fiber length and matrix temperature on the fractional anisotropy of fiber orientation.
The Folgar-Tucker equation and the improved anisotropic rotary diffusion model, incorporating analytical solutions, are used for verification.
Results show that the method improves significantly upon standard Monte Carlo estimation, while also providing error guarantees.
These findings offer the first step toward a reliable and practical tool for optimizing fiber-reinforced polymer manufacturing processes in the future.
\end{abstract}


\maketitle

\section{Introduction}\label{}
The use of polymer parts has reached nearly every industrial and economical sector of today's society.
Therefore, the production of such parts has an enormous environmental and ecological impact, and an optimized, robust, and stable production with minimal uncertainty and waste is desirable.
Especially when discontinuous fiber-reinforced polymers (FRPs) are molded, the process involves uncertainties in resulting fiber orientation and, consequently, in the final part's mechanical and thermal properties.
Today, the most common methodology to optimize molding processes and molded FRP parts in the early stages of product development are numerical simulations, which include a molding simulation with model-based fiber orientation prediction.
Such molding simulations are typically computational fluid dynamic simulations, which are based on the finite element method \cite{Tamil2012, Meyer2020, Sommer2018, Favaloro2018} or the finite volume method \cite{Ospald2014, Wittemann2018, Wittemann2019, Wittemann2021, Wittemann2022}.
One key aspect is the viscosity and flow modeling of the material, which determine the most important results of these simulations: the general feasibility of the molding, pressure and temperature distribution, and information about the resulting fiber orientation distribution.
However, the simulation approaches are purely deterministic and represent only one specific realization of possibly uncertain process and material parameters.

Uncertainty quantification (UQ) provides a framework for analyzing how randomness influences system behavior.
One of the most basic approaches to assessing the effects of the forward propagation of uncertainties is the Monte Carlo method, which is based on repeated random sampling and has been used for various different applications due to its non-intrusive nature.
However, this method can become inefficient and computationally prohibitive when applied to complex simulations.
Multilevel Monte Carlo (MLMC) methods \cite{Heinrich2001, Giles2008, Giles2015} offer a more efficient alternative by exploiting simulations of varying accuracy.
Originally, MLMC methods were developed to efficiently approximate expected values.
However, in safety-critical domains (e.g., aerospace or medical), where FRPs are widely used, it is necessary to quantify uncertainties beyond their average performance (i.e., expected value).
For instance, recently advanced MLMC methods for estimating central moments of arbitrary order have been introduced \cite{Bierig2015, Krumscheid2020}.
To ensure further reliability, it is important to estimate the likelihood of highly improbable events that occur at the extreme ends of a distribution, as these tail risks are critical for assessing potential extreme impacts.
Such risks are often quantified by metrics like quantiles for prescribed critical probabilities or failure probabilities.
Developments in this direction utilize nested simulation \cite{Giles2019}, adaptive MLMC methods \cite{HajiAli2022}, or approximate parametric expectations of suitable functions in order to subsequently obtain such risk quantities by post-processing \cite{Krumscheid2018, AyoulGuilmard2023}.

In this study, a different route is taken, focusing on a method to approximate multiple risk factors simultaneously.
For this, in the first step, the uncertain system output, or response, is approximated by a suitable, multivariate polynomial of a finite number of underlying parametric uncertainties.
Subsequently, this polynomial surrogate can be evaluated at a low computational cost, enabling the efficient estimation of several statistical quantities.
Once this surrogate is available, it eliminates the need to run a complex computational model (e.g., fiber orientation simulation), which is often the most computationally expensive step.
However, it raises the critical question of whether the simplified surrogate model is truly accurate and reliable.
As we will show, the accuracy of estimations of statistics derived from the surrogate model is directly connected to the accuracy of the surrogate.
In particular, we present novel upper bounds for the corresponding errors based on the surrogate approximation error.
This is done to ensure reliable UQ of statistics that are part of safety-critical application areas.
In order to construct a surrogate with a certain accuracy at minimum cost, we employ a hierarchical polynomial regression method \cite{HajiAli2020}, where multilevel techniques are applied to the optimal least squares operator \cite{Cohen2017}, rather than to the expectation operator, as was originally proposed.
These polynomial surrogates are then used for the estimation of aforementioned statistics of a quantity of interest, in particular the expected value, standard deviation, cumulative distribution function, and quantiles.

The rest of this work is structured as follows.
In Section \ref{sec:process_modeling}, we introduce the models and methods that form the basis of process modeling in the numerical experiments presented later.
Section \ref{sec:uq} outlines the multilevel polynomial regression approach and discusses the error bounds for estimating statistics using these polynomial surrogates.
Finally, Section \ref{sec:experiments} presents numerical experiments to verify these theoretical error bounds.
A list of symbols, divided into fiber orientation-related and mathematical symbols, is provided immediately after the conclusion Section \ref{sec:conclusion}.

\section{Process modeling}\label{sec:process_modeling}
In this section, we present the models and methods that form the basis for the numerical experiments in Section \ref{sec:experiments}.
The focus is on the modeling of discontinuous fiber-reinforced polymers, covering both the fluid mechanics and fiber orientation models essential for the simulations.
The forward uncertainty quantification methodology, which will provide a framework for systematically assessing uncertainties in these models, will be introduced in the subsequent Section \ref{sec:uq}.

\subsection{Fluid mechanics}\label{sec:fluid_dynamics}
Throughout this study, we consider the Hagen-Poiseuille equation \cite{Drazin2006} in a rectangular channel of width $w$ and height $h$, for which the solution of the velocity $v$ is given by
\begin{equation}\label{eq:hp}
    v(x, y) = \frac{4 P h^2}{\eta \pi^3} \sum_{i=1}^\infty \frac{1}{k_i^3} \left(1 - \frac{\cosh (\beta_i (x - \frac{w}{2}))}{\cosh (\beta_i \frac{w}{2})}\right) \sin (\beta_i y),
\end{equation}
for $(x, y) \in [0, w] \times [0, h]$, with $k_i = 2i-1$ and $\beta_i = \pi / h \, k_i$.
Here, $P$ is assumed to be a constant pressure gradient and $\eta$ the constant viscosity.

\subsection{Viscosity modeling}
One of the parameters that we will model in the numerical experiments with inherent uncertainty is the temperature of the material.
The temperature, in turn, affects the viscosity $\eta$, which subsequently impacts underlying fluid mechanics, as described, for instance, by Eq. \eqref{eq:hp}.
Viscosity modeling is performed using the Cross-WLF model \cite{Cross1965}, which, for temperatures $T>T^*$, where $T^*$ is the glass transition temperature, reads
\begin{equation}\label{eq:visc}
\eta(T) = \frac{\eta_0 (T)}{1 + \left(\frac{\eta_0 (T)\dot{\gamma}}{\tau^*} \right)^{1-n}},
\end{equation}
with a material specific parameter $n$, critical shear stress indicator $\tau^*$, shear rate $\dot{\gamma}$, and zero-shear viscosity
\begin{equation}\label{eq:zero_visc}
\eta_0(T) = D \exp \left(- \frac{A_1 (T - T^*)}{A_2 + (T-T^*)}\right),
\end{equation}
where $D, A_1, A_2$ are material-dependent, data-fitted coefficients.
Such data-fitted parameters are prone to measurement errors and could thus also be modeled with uncertainty.
For simplicity, we neglect shear thinning and assume a constant shear rate of $\dot{\gamma} = 1 s^{-1}$.
For the numerical experiments of this study, we will use existing data corresponding to a 30 wt\% glass fiber filled material (considered parameters are provided in Table \ref{tbl:basf} in Appendix \ref{apx:params}).

\subsection{Fiber orientation modeling}\label{sec:fiber_orientation}
In order to investigate the influence of various uncertain parameters on fiber orientation, the corresponding models are now presented. 
Instead of tracking the orientation of a large number of fibers separately, a homogenization procedure is performed according to \citet{Advani1987} using fiber orientation tensors $\mathbf{A} \in \mathcal{S}_+$, where
\begin{equation}\label{eq:S_+}
    \mathcal{S}_+ = \bigl\{ \mathbf{A} \in \mathbb{R}^{3 \times 3} : \mathbf{A} = \mathbf{A}^\top, \quad x^\top \mathbf{A} x \geq 0 \,\,\, \forall x \in \mathbb{R}^3, \quad \text{tr}(\mathbf{A}) = 1 \bigr\}
\end{equation}
denotes the space of symmetric and positive semi-definite $3 \times 3$ matrices with trace one.
The evolution of the fiber orientation is modeled using a fiber orientation model (FOM).
The most original model for the evolution of fiber orientation is the so-called Jeffery equation \cite{Jeffery1922} with its tensorial formulation
\begin{equation}\label{eq:jeffery}
    \dot{\mathbf{A}} = \mathbf{W} \mathbf{A} - \mathbf{A} \mathbf{W} + \xi \left ( \mathbf{D}\mathbf{A} + \mathbf{A}\mathbf{D} - 2 \mathbf{D} : \mathbb{A}\right),
\end{equation}
where $\mathbf{D} = (\mathbf{L} + \mathbf{L}^\top) / 2$ is the rate-of-deformation and $\mathbf{W} = (\mathbf{L} - \mathbf{L}^\top) / 2$ the vorticity tensor, computed using the velocity gradient $\mathbf{L} \in \R^{3 \times 3}$ of some fluid in motion.
Two other important variables in Eq. \eqref{eq:jeffery} are the particle shape factor $\xi$ and the fourth order fiber orientation tensor $\mathbb{A} \in \R^{3\times 3\times 3 \times 3}$.
Typically, the latter is approximated $\mathbb{A} \approx f(\mathbf{A})$ using a closure function $f$ and the second-order orientation tensor $\mathbf{A}$.
The Folgar-Tucker equation (FTE) \cite{Folgar1984} with its tensorial formulation
\begin{equation}\label{eq:fte}
    \dot{\mathbf{A}} = \mathbf{W} \mathbf{A} - \mathbf{A} \mathbf{W} + \xi \left ( \mathbf{D}\mathbf{A} + \mathbf{A}\mathbf{D} - 2 \mathbf{D}: \mathbb{A} \right) + 2C_I \dot{\gamma} (\mathbf{I} - 3\mathbf{A})
\end{equation}
extends Jeffery's equation by also accounting for rotary diffusion due to fiber-fiber interactions, controlled through the fiber interaction coefficient $C_I > 0$. The shear rate is given by $\dot{\gamma} = (2 \text{tr}(\mathbf{D}^2))^{1/2}$.

Another FOM, particularly suited for long fiber thermoplastics, is the anisotropic rotary diffusion model (ARD) developed by \citet{Phelps2009}, which reads
\begin{equation}\label{eq:ard}
    \dot{\mathbf{A}} = \mathbf{W} \mathbf{A} - \mathbf{A} \mathbf{W} + \xi \left ( \mathbf{D}\mathbf{A} + \mathbf{A}\mathbf{D} - 2 \mathbf{D} : \mathbb{A}\right) + \dot{\gamma} \left[2 \mathbf{C} - 2 \text{tr}(\mathbf{C}) \mathbf{A} - 5 (\mathbf{C}\mathbf{A} + \mathbf{A}\mathbf{C}) + 10 \mathbb{A}:\mathbf{C}\right].
\end{equation}
The scalar interaction coefficient $C_I$ from Eq. \eqref{eq:fte} is replaced by a rotary diffusion tensor $\mathbf{C} \in \R^{3\times3}$.
In particular, choosing
\begin{equation}\label{eq:iard_diff_tensor}
    \mathbf{C} = C_I [\mathbf{I} - 4 C_M (\mathbf{D} / \dot{\gamma})^2]
\end{equation}
with an anisotropy controlling parameter $C_M \in [0, 1]$ yields the improved ARD (iARD) model \cite{Tseng2016}.
Analytical solutions for both the FTE and the ARD model with hybrid closure were derived by \citet{Winters2022} in general form.
Detailed formulas used for the numerical experiments of this study are included in Appendix \ref{apx:analytical_sols} for reference.

\section{Uncertainty Quantification}\label{sec:uq}
Uncertainty quantification in predicting fiber orientation distributions is essential for understanding the variability in process outcomes and their impact on the mechanical properties of manufactured parts.
In this section, we present a general forward UQ framework based on polynomial regression for the assessment of the propagation of such uncertainties.
More precisely, we introduce multilevel polynomial surrogates in Section \ref{sec:ml_l2}, demonstrate how they can be used a-posteriori to estimate statistical quantities in Section \ref{sec:comp_stats}, and provide novel error bounds for these estimations in Section \ref{sec:ml_l2_error}.

\subsection{Multilevel polynomial surrogates}\label{sec:ml_l2}
A common challenge in uncertainty quantification is dealing with solutions to differential equations influenced by randomness.
For instance, this could be an FOM from Section \ref{sec:fiber_orientation}, where the fiber interaction coefficient is experimentally determined and therefore subject to uncertainty.
Denote by $u(\cdot, \omega)$ the solution to such a differential equation given a random realization of parameters $\omega \in \Omega \subseteq \R^d$ with $\omega \sim \mu$, where $\mu$ is a probability measure on $\Omega$.
In addition to the actual solution, attention is primarily centered on a specific Quantity of Interest (QoI) derived from the solution, denoted as $q(u) \in \R$.
For instance, such a QoI could represent the solution's value at specific points in time or space, or its spatial average.
The following method aims to approximate the impact of random parameters on the QoI, represented by the mapping
\begin{align}\label{eq:general_response_surface}
\begin{split}
    Q: \Omega &\rightarrow \R, \\
    \omega &\mapsto Q(\omega) := q(u(\cdot, \omega)),
\end{split}
\end{align}
which is also referred to as the \emph{response surface}.
Note that evaluating this response necessitates solving a potentially complicated differential equation, for which an explicit solution typically does not exist.
Consequently, numerical methods must be employed, which often involve significant computational costs and induce discretization errors.
Therefore, one goal of UQ is to develop surrogate models that emulate the response surface and thus avoid the computational effort of solving the underlying differential equation.

\subsubsection{Polynomial surrogates}
In this study, we follow \citet{HajiAli2020} to construct and further use a (multivariate) polynomial surrogate
\begin{equation}\label{eq:surrogate}
\hat{Q}(\omega) = \sum_{\lambda \in \Lambda} q_\lambda \, \omega^\lambda 
\end{equation}
to approximate the unknown response surface $Q$.
Here, $\Lambda \subseteq \N_0^d$ denotes the index set and $q_\lambda \in \R$ the coefficients.
Index sets offer a convenient representation of multivariate polynomials by using multi-indices to extend naturally through tensorization.
For example, given an index $\lambda = (\lambda_1, \ldots, \lambda_d) \in \Lambda$, the monomial appearing in Eq. \eqref{eq:surrogate} is defined by
\begin{equation}
    \omega^\lambda := \prod_{j=1}^d \omega_j^{\lambda_j}.
\end{equation}
Depending on the regularity of the response surface $Q$, different index sets can lead to significant differences in accuracy when computing a polynomial approximation \cite{Trefethen2019}.
The most natural index set is the tensor product (TP) set, which, for $m \in \N_0$, is given by indices of degree at most $m$ in each dimension, that is
\begin{equation}\label{eq:tensor_product}
    \Lambda_{TP}(m) := \{\lambda \in \N^d_0: \max_{i =1,\ldots, d} |\lambda_i| \leq m\}.
\end{equation}
Once an accurate polynomial surrogate with $\hat{Q} \approx Q$ is available, 
it offers the significant advantage of being computationally cheap to evaluate.
This makes it an ideal choice for post-processing tasks that may require a large number of samples, such as Monte Carlo type methods.
The estimation of various statistical parameters of the QoI through the use of a polynomial approximation $\hat{Q}$ is discussed in Sections \ref{sec:comp_stats} and \ref{sec:ml_l2_error}.
Next, we present the construction of the polynomial surrogate employed in this study.

\subsubsection{Weighted least squares}\label{sec:weighted_least_squares}
The process of approximating a function $Q$ through evaluations $\{Q(\omega_1), \ldots, Q(\omega_N)\}$ in so-called collocation points $\omega_i$ is well-researched.
One approach involves using weighted discrete least squares regression \cite{Cohen2017}.
For this, let $\mu$ be a probability measure on $\Omega$ and
\begin{equation}
    L^2_\mu(\Omega) := \left\{ f: \Omega \rightarrow \R : \int_\Omega |f(\omega)|^2 \,\mu(d\omega) < \infty \right\}
\end{equation}
the space of functions with finite $L^2_\mu$ norm, where
\begin{equation}\label{eq:l2_norm}
    \|f\|_{L^2_\mu(\Omega)} = \left(\int_\Omega |f(\omega)|^2 \,\mu(d\omega)\right)^\frac{1}{2}.
\end{equation}
For notational brevity, in the following we will use $\|\cdot\| := \|\cdot\|_{L^2_\mu(\Omega)}$.
In this study, $\mu$ is considered to be the Lebesgue measure, which corresponds to a uniform distribution on bounded $\Omega$ and represents the orthogonality weighting for the Legendre polynomials.
Other distributions can be converted to the uniform distribution by inverse transforming.
Given an element $Q \in L^2_\mu(\Omega)$ and a polynomial subspace $V \subset L^2_\mu(\Omega)$, the goal is to find its best approximation
\begin{equation}\label{eq:l2_best_approx}
    \argmin_{v \in V} \|Q - v\|^2 =: \Pi_V Q,
\end{equation}
that is, its orthogonal projection $\Pi_V Q \in V$.
Given independent and identically distributed Monte Carlo samples $\omega_1, \ldots, \omega_N \in \Omega$ with $\omega_i \sim \mu$, the best approximation can be approximated by minimizing the discrete least squares seminorm
\begin{equation}
    \frac{1}{N} \sum_{i=1}^N |Q(\omega_i) - v(\omega_i)|^2.
\end{equation}
One technique for reducing the variance in such Monte Carlo estimations is the use of \textit{importance sampling} \cite{Kloek1978}.
Rather than sampling from the underlying orthogonal measure $\mu$, samples are drawn from alternative distributions that better emphasize regions of the domain that are critical for the desired quantity.
The corresponding discrete \textit{weighted} least squares approximation of Eq. \eqref{eq:l2_best_approx} is given by
\begin{equation}\label{eq:lsq}
    \argmin_{v \in V} \|Q - v\|^2_N =: \Pi_V^N Q,
\end{equation}
where
\begin{equation}\label{eq:semi_norm}
    \|f\|_N^2 := \frac{1}{N} \sum_{i=1}^N \left(\frac{d\mu}{d\nu}\right)(\omega_i) \,|f(\omega_i)|^2
\end{equation}
denotes the weighted discrete squared semi-norm based on an independent and identically distributed sample $\{\omega_i\}_{i=1}^N \subset \Omega$.
Again, the key feature here is the ability to draw these samples from a different measure $\omega_i \sim \nu$, with $\nu \ll \mu$, rather than from the underlying measure $\mu$.
The weights in Eq. \eqref{eq:semi_norm} defined by the (inverse of the) Radon–Nikodym derivative $d\nu/d\mu$ correct for the introduced bias, ensuring that the squared semi-norm remains an unbiased Monte Carlo estimation for the underlying squared $L^2$ norm.
This way, it is expected that the discrete weighted least squares fit, Eq. \eqref{eq:lsq}, provides a good estimation to the best approximation, Eq. \eqref{eq:l2_best_approx}.

There exists an optimal sampling distribution choice $\nu = \nu(\mu, V)$, which depends on the underlying measure $\mu$ and on the polynomial subspace $V$, and can be constructed accordingly.
Using this distribution, the number of samples needed to achieve an $L^2$ error comparable to the best approximation, is of order $N = \mathcal{O}(m \log m)$, where $m$ is the dimension of $V$ (see \cite{Cohen2017} for details).

\subsubsection{Multilevel weighted least squares}\label{sec:ml_weighted_least_squares}
Recall that evaluating the response surface, Eq. \eqref{eq:general_response_surface}, involves solving a differential equation, for which an explicit solution $u(\cdot, \omega)$ rarely exists.
Therefore, numerical solvers must be used, which provide only approximated solutions $u_n$ for a certain discretization parameter $n \in \N$, and thus approximate response surfaces
\begin{equation}
    \omega \mapsto Q_n(\omega) := q(u_n(\cdot, \omega)).
\end{equation}
For example, $Q_n$ could represent the mapping obtained by applying the QoI to a numerical solution $u_n$, which was computed using a finite difference method with $n \in \N$ discretization points.
Other discretization schemes such as the finite volume or finite element method can be used too, of course.
We therefore assume that $Q_n \rightarrow Q$ for $n \rightarrow \infty$ in an appropriate sense, with increasing computational costs of evaluating $Q_n$ as $n \rightarrow \infty$.

Multilevel techniques leverage simulations of varying discretization levels to obtain a more efficient method when combined.
This has been successfully applied \cite{HajiAli2020} to polynomial weighted discrete least squares, which will be briefly outlined in the following.
For a given number of levels $L \in \N$, a sequence of increasing discretization parameters $n_0, n_1, \ldots, n_L \in \N$ is selected.
To enhance and to actually observe the efficiency of multilevel methods, these discretization parameters are often modeled exponentially.
For example, they can be set as $n_l = 2^l$, doubling the number of discretization points with each level $l \in \{0, \ldots, L\}$.
To introduce the multilevel version of Eq. \eqref{eq:lsq}, one first expands the most accurate simulation $Q \approx Q_{n_L}$ using the telescoping sum
\begin{equation}\label{eq:telescoping_sum}
    Q_L = Q_0 + \sum_{l=1}^L Q_l - Q_{l-1},
\end{equation}
where we abbreviated $Q_l := Q_{n_l}$ for $l \in \{0, \ldots, L\}$.
This sum can be interpreted as first approximating $Q$ using a coarse and cheap approximation $Q_0$, and then correcting the error $Q - Q_0$ through difference terms.
The idea now is to approximate each term in Eq. \eqref{eq:telescoping_sum} independently using polynomial weighted least squares, Eq. \eqref{eq:lsq}.
Because of $Q_n \rightarrow Q$ and therefore also $Q_n - Q_{n-1} \rightarrow 0$ as $n \rightarrow \infty$, the correction terms are of decreasing order and can therefore be represented by shrinking polynomial spaces.
We select a sequence of increasing polynomial space parameters $m_0, m_1, \ldots, m_L \in \N$, corresponding to polynomial spaces $V_{m_l} \subset L^2_\mu(\Omega)$, and define the multilevel weighted discrete least squares method
\begin{equation}\label{eq:ml_lsq}
    \hat{Q} := \Pi_{V_L} Q_0 + \sum_{l=1}^L \Pi_{V_{L-l}} \left(Q_l - Q_{l-1}\right),
\end{equation}
where we again abbreviated $V_l := V_{m_l}$ for $l \in \{0, \ldots, L\}$.
Such polynomial space parameters can represent, for instance, the maximum polynomial degree, which are employed to define tensor product index sets, Eq. \eqref{eq:tensor_product}.
The exact choice of the sequences $n_l$ or $m_k$ depends exponentially on the convergence and cost rate of the underlying numerical solver, as well as the growth rate and approximability of the underlying polynomial spaces.
For example, the $n_l$ are modeled to grow faster when the underlying numerical solver has lower costs and slower convergence.
In contrast, the $m_k$ are modeled to grow slower when the polynomial spaces grow more quickly or have poorer polynomial approximability (see \cite{HajiAli2020} for the definition of $n_l$ and $m_k$).

Based on a user-specified error tolerance $\epsilon > 0$, the multilevel method is designed to construct a polynomial $\hat{Q}_\epsilon$, which, due to the optimal sampling strategy described in Section \ref{sec:weighted_least_squares}, is a random variable, that satisfies
\begin{equation}
    \| Q - \hat{Q}_\epsilon\| \leq \epsilon
\end{equation}
with probability larger than $1 - \epsilon^{\log \log (\epsilon^{-1})}$ (\cite{HajiAli2020}, Theorem 4.3).
Furthermore, the asymptotic computational costs for computing $\hat{Q}_\epsilon$ are never greater than those of a single-level counterpart, Eq. \eqref{eq:lsq}, which uses evaluations of a single numerical solver discretization, i.e., computing $\Pi_V Q_n$ for some discretization parameter $n \in \N$ and fixed finite-dimensional subspace $V \subset L^2_\mu(\Omega)$.
For cases where the convergence rates, evaluation costs, or polynomial approximability of the function sequences $Q_n$ are not or only partially known, there is an adaptive analog to Eq. \eqref{eq:ml_lsq}.
This adaptive algorithm, which is also used in this work, iteratively constructs an index set and a sequence of functions $Q_n$ to optimally balance computational effort with the resulting $L^2$ error (see \cite{HajiAli2020} for details).

\subsection{Computation of statistical quantities}\label{sec:comp_stats}
Access to an accurate and easy to evaluate surrogate for the response surface enables the efficient computation of statistics of the QoI, e.g., using the Monte Carlo method.
This could include statistical moments such as the expected value or the variance, as well as risk measures like failure probabilities or quantiles.
In this section, the computation of such statistics via a polynomial surrogate shall be explained.
Let $V = \text{span}\{\omega^\lambda : \lambda \in \Lambda\} \subset L^2_\mu(\Omega)$
be a polynomial subspace spanned by monomials given by an index set $\Lambda \subset \N_0^d$.
Further, let $\{P_\lambda\}_{\lambda \in \Lambda}$ be an $L^2_\mu(\Omega)$ orthonormal basis of $V$, which means that for the expected value of the product of two basis elements, we have
\begin{equation}
    \E[P_{\lambda}P_{\xi}] = \int_\Omega P_\lambda(\omega) P_\xi(\omega) \,\mu(d \omega) = \delta_{\lambda \xi} \quad \forall \lambda, \xi \in \Lambda,
\end{equation}
where
\begin{equation}
    \delta_{\lambda \xi} = 
    \begin{cases} 
          1, & \text{if} \,\, \lambda = \xi \\
          0, & \text{else}
       \end{cases}
\end{equation}
denotes the Kronecker delta.
Then, for a polynomial
\begin{equation}
    V \ni \hat{Q} = \sum_{\lambda \in \Lambda} q_\lambda P_\lambda,
\end{equation}
some quantities can be obtained analytically by post-processing its coefficients.
For instance, by exploiting orthonormality, the mean is given by the first coefficient,
\begin{equation}
    \E[\hat{Q}] = q_\mathbf{0},
\end{equation}
and the variance by the sum of the squares of the remaining coefficients,
\begin{equation}
    \V[\hat{Q}] = \sum_{\lambda \in \Lambda \setminus \{\mathbf{0}\}} q_\lambda^2.
\end{equation}
Additional statistical metrics, such as the cumulative distribution function (CDF) and the $p$-quantile, must be estimated empirically.
This can be achieved using the Monte Carlo method, which will be further elaborated upon in the numerical experiments in Section \ref{sec:experiments}.

\subsection{Error bounds for statistical quantities}\label{sec:ml_l2_error}
Once statistical quantities are computed a-posteriori, it is essential to assess their accuracy.
Hereafter, we present error bounds on a few of such quantities by leveraging the ability to control $\| Q - \hat{Q}_\epsilon\| \leq \epsilon$ through the multilevel method, Eq. \eqref{eq:ml_lsq}.

Define $\hat{Q} := \hat{Q}_\epsilon$ for some $\epsilon > 0$.
We start by using Jensen's inequality to bound the error of the expected values by
\begin{equation}\label{eq:jensen}
    |\E\left[Q\right] - \E[\hat{Q}]|^2 \leq \E\left[| Q - \hat{Q}|^2\right] = \| Q - \hat{Q}\|^2.
\end{equation}
To bound the error of the standard deviations, we use the reverse triangle inequality to obtain
\begin{equation}
    |\V[Q]^{1/2} - \V[\hat{Q}]^{1/2}| \leq \|Q-\hat{Q}\|.
\end{equation}
A detailed derivation of this and the following bounds can be found in Appendix \ref{apx:error_bounds}.
Denote by $F_X := \mathbb{P}(X \leq \cdot)$ the CDF of a random variable $X$.
A uniform bound for the error between the CDF of $Q$ and its polynomial approximation $\hat{Q}$ is given by
\begin{equation}\label{eq:cdf_err}
    \| F_Q - F_{\hat{Q}}\|_\infty \leq 3 \left(\| f_Q \|_\infty \| Q-\hat{Q}\|\right)^{\frac{2}{3}},
\end{equation}
where $f_Q$ is the PDF of $Q$ and $\|f\|_\infty := \sup_{x \in \R} |f(x)|$ denotes the supremum norm.
Finally, for the $p$-quantile with probability $p \in (0, 1)$ of a random variable $X$, that is $q_X := \inf \{q \in \R: F_X(q) \geq p\}$,
we have
\begin{equation}\label{eq:quantile_bound}
    |q_Q - q_{\hat{Q}}| \leq 3 f_Q(q)^{-1} \left(\| f_Q \|_\infty \| Q-\hat{Q}\|\right)^{\frac{2}{3}},
\end{equation}
where $q \in \text{conv}(q_Q, q_{\hat{Q}})$ depends on the probability $p$.
Thus, the derived error bounds, Ineqs. \eqref{eq:jensen} - \eqref{eq:quantile_bound}, enable us to rigorously quantify the error in the statistics of the QoI when a polynomial surrogate $\hat{Q}$ satisfies $| Q - \hat{Q}_\epsilon| \leq \epsilon$ for a given error tolerance $\epsilon > 0$.

\subsection{Global error bounds for statistical quantities}\label{sec:global_surrogates}
A QoI often represents a specific point in time or space of a model's solution, or it may be derived using a specific set of parameter configurations.
Ideally, the investigation of a QoI would not be limited to a single configuration or specific point but would extend to multiple constellations to detect potential anomalies or to assess uncertainties across different points in time or space.
To illustrate, consider a molding process where we are interested in the fiber orientation of an FRP.
Rather than approximating the QoI, such as the fiber orientation, at one specific material point, we aim to develop an approximation that spans the entire geometry of the part, denoted as $\mathcal{X} \subset \R^p$.
One possible approach to tackle this would be to construct a polynomial surrogate for each point in the geometry.
By inter- and extrapolating these polynomials (or their statistics), an approximation could be obtained across the entire parts geometry.
However, the number of grid points required for an accurate approximation would be high, fitting a polynomial to each grid point is computationally intensive, and the established error bounds would only apply to those specific grid points.

\begin{figure*}[!ht]
\begin{subfigure}{.5\textwidth}
\centering
\begin{tikzpicture}
\node[align=center] at (-3.5,0) {\textbf{Uncertainties}\\$\omega \in \Omega$};
\node[align=center] at (0,1.3) {Fixed $x \in \mathcal{X}$};
\draw[dashed,->] (0,1.1) -- (0,.6);
\node[draw,,align=center] at (0,0) {\textbf{Model} \\ $u(\cdot, \omega)$};
\draw[->] (-2.2,0) -- (-1.2,0);
\draw[->] (1.2,0) -- (2.2,0);
\node[align=center] at (3,0) {\textbf{QoI}\\$Q(\omega)$};
\end{tikzpicture}
\vspace{.25cm}
\caption{Fixed parameter $x \in \mathcal{X}$}
\label{fig:diagram_a}
\end{subfigure}%
\begin{subfigure}{.5\textwidth}
\centering
\begin{tikzpicture}
\node[align=center] at (-3.5,.6) {\textbf{Uncertainties}\\$\omega \in \Omega$};
\node[align=center] at (-3.5,-.6) {\textbf{Parameter}\\$x \in \mathcal{X}$};
\node[draw,,align=center] at (0,0) {\textbf{Model} \\ $u(\cdot, \omega)$};
\draw[->] (-2.2,.6) -- (-1.2,.2);
\draw[->] (-2.2,-.6) -- (-1.2,-.2);
\draw[->] (1.2,0) -- (2.2,0);
\node[align=center] at (3,0) {\textbf{QoI}\\$Q(\omega, x)$};
\end{tikzpicture}
\vspace{.25cm}
\caption{Arbitrary parameter $x \in \mathcal{X}$}
\label{fig:diagram_b}
\end{subfigure}
\caption{Illustration of model input and output relationships: (a) fixed parameter case $x \in \mathcal{X}$; (b) parameter augmented case, where both uncertainties $\omega \in \Omega$ and parameters $x \in \mathcal{X}$ are treated as random inputs.}
\label{fig:diagram}
\end{figure*}
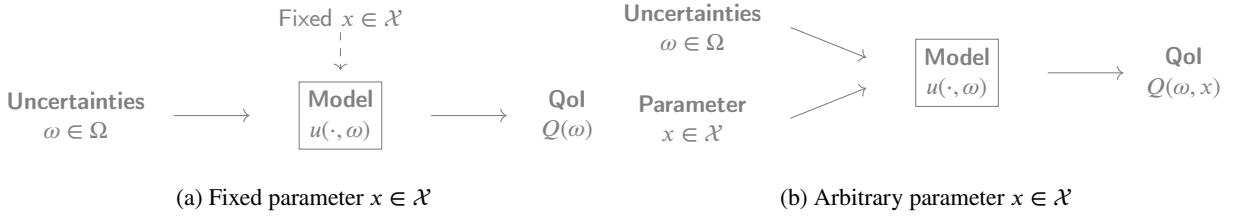

In this subsection, we remove this locality restriction by including the parameters $x \in \mathcal{X}$ to the group of uncertain parameters.
Similar to the previously considered uncertainties $\omega \in \Omega$, they usually enter the model evaluation at a certain stage and affect subsequent computations.
Combining the parameters $x \in \mathcal{X}$ and the uncertainties $\omega \in \Omega$ allows for flexibility in the model without committing to any particular configuration of $x$, that needs to be fixed a-priori.
This is visualized in Figure \ref{fig:diagram}, which compares the setting with a fixed parameter configuration $x \in \mathcal{X}$ to the global approach, where the response surface $Q$ incorporates the parameter $x$.
The motivation of this approach is that local surrogate models could be obtained by partially evaluating a global surrogate.
To be more precise, consider, similar to Eq. \eqref{eq:general_response_surface}, the response surface
\begin{equation}\label{eq:global_response}
    Q: \Omega \times \mathcal{X} \rightarrow \mathbb{R},
\end{equation}
defined on the domain of actual uncertainties $\Omega \subset \R^d$ and further parameters $\mathcal{X} \subset \R^p$.
Let $\mu_\omega$ and $\mu_x$ be probability measures on $\Omega$ and $\mathcal{X}$, respectively. Denote by $\mu = \mu_{\omega} \otimes \mu_{x}$ the product measure on $\Omega \times \mathcal{X}$, and we retain the assumption of the Lebesgue measure.
As before, a polynomial approximation $\hat{Q}: \Omega \times \mathcal{X} \rightarrow \R$ of this response surface can be constructed by using the multilevel estimator, Eq. \eqref{eq:ml_lsq}.
The question that arises is whether the partial evaluation at a specific point $x \in \mathcal{X}$, that is
\begin{equation}
    \hat{Q}\left(\cdot, x\right): \Omega \rightarrow \R,
\end{equation}
represents an accurate approximation of a response surface that was constructed solely using collocation points $\{\omega_i\} \subset \Omega$ with the parameter $x \in \mathcal{X}$ held fixed.
This approach offers the advantage of requiring only a single, higher-dimensional polynomial, which involves more variables (increasing from $d$ to $d+p$ parameters), while providing grid-free access to local surrogates.
In the following, we explore if and how the error observations from the beginning of Section \ref{sec:ml_l2_error} carry over.
Let $\hat{Q}$ be a polynomial approximation of $Q$, Eq. $\eqref{eq:global_response}$.
For any $x \in \mathcal{X}$, Lemma \ref{lemma:partial_eval} from Appendix \ref{apx:partial_eval} states that
\begin{equation}\label{eq:bound_partial_eval}
    \|Q(\cdot, x) - \hat{Q}(\cdot, x)\|^2_{L^2_{\mu_\omega}} < C_1 \left(C_2 + \|Q - \hat{Q}\|^2_{L^2_{\mu}}\right),
\end{equation}
where $C_1$ depends on $x \in \mathcal{X}$ and $C_2$ on the Legendre coefficients of $Q - \hat{Q}$.
Here, as in Eq. \eqref{eq:l2_norm}, $\|\cdot\|_{L^2_{\mu_\omega}}$, $\|\cdot\|_{L^2_{\mu_x}}$, and $\|\cdot\|_{L^2_{\mu}}$ denote the $L^2$ norms on $\Omega, \mathcal{X}$, and $\Omega \times \mathcal{X}$, respectively, with respect to measures $\mu_x, \mu_\omega$, and $\mu$.
This inequality shows that the $L^2$ error control between the partial evaluation and its polynomial approximation carries over only to some extent.
If
\begin{equation}
    \|Q - \hat{Q}\|_{L^2_\mu}^2 \leq \epsilon
\end{equation}
for some prescribed tolerance $\epsilon > 0$, we can only ensure an error within $C_1 (C_2 + \epsilon)$.
This outcome is expected, as $L^2$ convergence does not imply pointwise convergence in general.
In the simpler case where $Q - \hat{Q}$ is a polynomial, full error control for the partial evaluations is achieved (see Remark \ref{rm:polynomial} in Appendix \ref{apx:partial_eval}).

Instead, we bound the quadratic error across the set $\mathcal{X}$ for the statistics directly.
For instance, if we define the expected value functions $m(x) := \E[Q(\cdot, x)]$ and $\hat{m}(x) := \E[\hat{Q}(\cdot, x)]$, it holds that
\begin{equation}
    \|m - \hat{m}\|_{L^2_{\mu_x}} \leq \|Q - \hat{Q}\|_{L^2_\mu}.
\end{equation}
The derivation of this and the subsequent inequalities is provided in detail in Appendix \ref{apx:error_bounds_ext}.
Similar, for the error in standard deviation functions $s(x) := \V[Q(\cdot, x)]^{1/2}$ and $\hat{s}(x) := \V[\hat{Q}(\cdot, x)]^{1/2}$, one obtains
\begin{equation}
    \|s - \hat{s}\|_{L^2_{\mu_x}} \leq \|Q - \hat{Q}\|_{L^2_\mu}.
\end{equation}
For the CDF, defining $F_z(x) := \mathbb{P}(Q(\cdot, x) \leq z)$ and $\hat{F}_z(x) := \mathbb{P}(\hat{Q}(\cdot, x) \leq z)$ for $z \in \R$, we have
\begin{equation}
    \sup_{z \in \R} \|F_z - \hat{F}_z\|^2_{L^2_{\mu_x}} \leq 27 \left(\| f_Q \|_\infty \|Q-\hat{Q}\|_{L^2_\mu}\right)^{\frac{2}{3}}.
\end{equation}
Lastly, for the $p$-quantiles $q(x) := q_{Q(\cdot, x)}$ and $\hat{q}(x) := q_{\hat{Q}(\cdot, x)}$, we have
\begin{equation}
    \|q-\hat{q}\|_{L^2_{\mu_x}} \leq 3 c^{-1} \left( \| f_Q \|_{\infty} \|Q - \hat{Q}\|_{L^2_{\mu}}\right)^{\frac{2}{3}}.
\end{equation}

\section{Numerical experiments}\label{sec:experiments}
In this section, we illustrate the theoretical bounds from Section \ref{sec:ml_l2_error} and validate the convergence rates using an example from molding simulation and fiber orientation modeling with uncertain material parameters.
The experiment is twofold, with a focus on a single material point in the first part, Section \ref{sec:exp_1}, and an extension of the analysis to the entire component geometry in Section \ref{sec:exp_2}.
All parameters relevant to the numerical experiment are listed in Tables \ref{tbl:basf} and \ref{tbl:params} in Appendix \ref{apx:params}.
The fiber orientation models, closure approximations, and equivalent aspect ratios used in this study were computed using \texttt{fiberoripy} \cite{fiberoripy}.

We introduce uncertainties that influence the orientation of fibers, thereby affecting the mechanical properties of a fiber-reinforced polymer.
More specifically, the impact of uncertainty in material temperature and fiber length on fiber orientation distribution is investigated by quantifying the fractional anisotropy.
Serving as our QoI, fractional anisotropy is a critical metric for assessing alignment within an orientation tensor (see Eq. \eqref{eq:S_+}).
It is defined by $F: \mathcal{S}_+ \rightarrow [0, 1]$ with
\begin{equation}\label{eq:fa}
F(\mathbf{A}) = \sqrt{\frac{3}{2}} \frac{\| \bm{\lambda} - 1/3 \|_2}{\|\bm{\lambda}\|_2},
\end{equation}
where $\bm{\lambda} = (\lambda_1, \lambda_2, \lambda_3)$ are the eigenvalues of the matrix $\mathbf{A} \in \mathcal{S}_+$ and $\|\cdot\|_2$ denotes the Euclidean norm.
The difference in the numerator of Eq. \eqref{eq:fa} is to be understood componentwise.

\begin{figure}[h!]
    \centering
    \includegraphics[width=.5\linewidth]{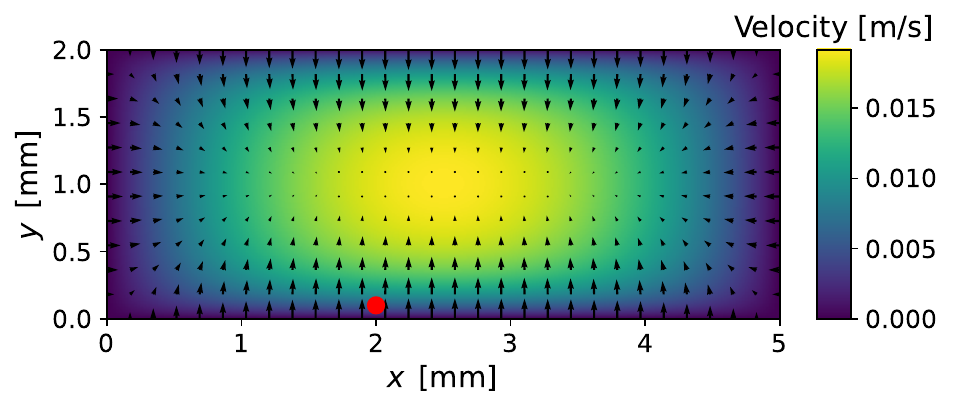}
    \caption{Hagen-Poiseuille flow $v(x, y)$, Eq. \eqref{eq:hp}, with viscosity $\eta = \eta(T)$ for $T=565$ K. The red dot marks the arbitrary selected point in the channel $(x_0, y_0) := (2.0 ,\, 0.1)$ mm, where fiber orientation is investigated. The arrows show the velocity gradient.}
    \label{fig:hp}
\end{figure}

The first material property to be modeled as uncertain is temperature $T \in \Omega_T \subset \R$, which directly influences the viscosity $\eta = \eta(T)$, described by the Cross-WLF model, as defined in Eqs. (\ref{eq:visc}, \ref{eq:zero_visc}).
The parameters used for this correspond to a 30 wt\% glass fiber filled material \cite{Moldflow2023} (see Table \ref{tbl:basf}).
The viscosity $\eta$, in turn, significantly affects fluid dynamics, which is modeled through the Hagen-Poiseuille flow described in Section \ref{sec:fluid_dynamics}.
For numerical computations, we truncate the sum in Eq. \eqref{eq:hp} after the first ten summands.
Figure \ref{fig:hp} depicts the Hagen-Poiseuille flow in a rectangular channel with width $w=5$ mm and height $h=2$ mm for a specific viscosity $\eta = \eta(T)$ at $T=565$ K.
Interpreting Eq. \eqref{eq:hp} as the flow in $z$-direction, the velocity gradient $\mathbf{L}$ in $(x, y) \in [0, w] \times [0, h]$, can be expressed as
\begin{equation}\label{eq:velocity_grad}
    \mathbf{L}(x, y) =
    \begin{pmatrix}
    0 & 0 & 0\\
    0 & 0 & 0\\
    \partial_x v(x, y) & \partial_y v(x, y) & 0
\end{pmatrix},
\end{equation}
which is used to compute the rate-of-deformation $\mathbf{D}$ and the vorticity tensor $\mathbf{W}$.

The second uncertainty, fiber length $L_f \in \Omega_{L_f} \subset \R$, on the other hand, alters the particle shape factor
\begin{equation}
    \xi = \frac{r_f^2 - 1}{r_f^2 + 1}
\end{equation}
through $r_f = g(L_f / D_f)$, the aspect ratio between fiber length and diameter.
Hereby, $g$ denotes the equivalent aspect ratio transformation for ellipsoids based on \citet{Zhang2011}.

\subsection{Fixed material point}\label{sec:exp_1}
In this first part, we select an arbitrary but fixed point $(x_0, y_0) := (2.0 ,\, 0.1)$ within the channel $[0, w] \times [0, h]$ (see Figure \ref{fig:hp}) and investigate the fractional anisotropy in this point only.
For this purpose, we interpret the Hagen-Poiseuille flow and the velocity gradient derived from it as a function of the temperature $T \in \Omega_T$ in $(x_0, y_0)$.

\begin{figure}[h!]
  \centering
   \includegraphics[width=.5\linewidth]{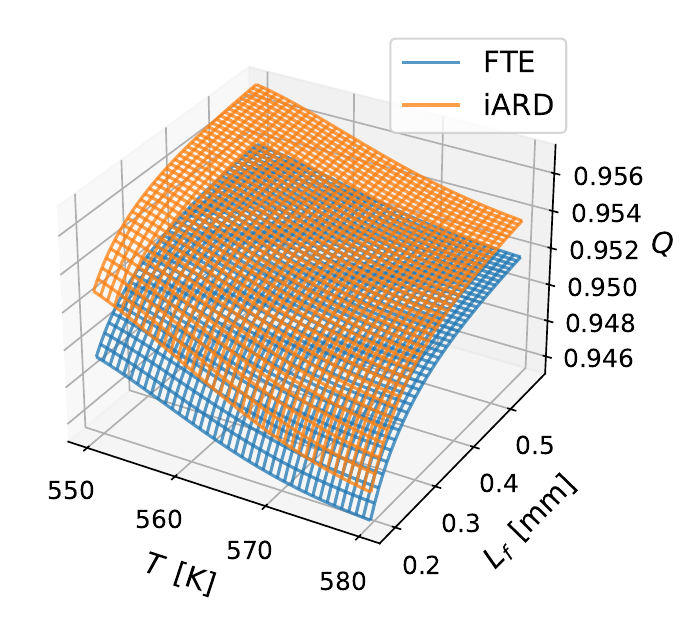}
    \caption{Response surface $Q = F\left(\mathbf{A}[t_\text{end}; \cdot]\right)$ showing the fractional anisotropy over the parameter domains $\Omega_T = [550, 580]$ K (temperature) and $\Omega_{L_f} = [0.38 \pm 50\%]$ mm (fiber length). Computations use the FTE and iARD model as FOM, respectively, with isotropic initial orientation tensor $\mathbf{A}(0) = \mathbf{I} / 3$.}
    \label{fig:response_surface}
\end{figure}

As mentioned in Section \ref{sec:fiber_orientation}, the evolution of the fiber orientation is modeled using an FOM, a matrix-valued differential equation
\begin{equation}\label{eq:FOM}
    \dot{\mathbf{A}} = M(\mathbf{A}; \omega), 
\end{equation}
with specifying right-hand side $M: \mathcal{S}_+ \times \Omega \rightarrow \mathcal{S}_+$.
Hereby, $\mathcal{S}_+$ represents the space of admissible orientation tensors, as defined in Eq. \eqref{eq:S_+}, while $\Omega \subseteq \R^d$ denotes the space of modeled uncertainties.
In this experiment, both temperature and fiber length have an impact on the fiber orientation distribution, forming the random realization of parameters
\begin{equation}
    \omega = (T, L_f) \in \Omega_T \times \Omega_{L_f} =: \Omega.
\end{equation}
We utilize both, the FTE, Eq. \eqref{eq:fte}, and iARD model, Eqs. (\ref{eq:ard}, \ref{eq:iard_diff_tensor}), as FOM in this numerical experiment.
The quadratic closure \cite{Doi1981} is employed for approximating the fourth-order orientation tensor.
The response surface to be approximated can be expressed as $Q: \Omega \rightarrow [0, 1]$ with
\begin{equation}\label{eq:response_surface}
    Q(\omega) = F\left(\mathbf{A}[t_\text{end}; \omega]\right),
\end{equation}
where $t \mapsto \mathbf{A}[t; \omega]$ represents the solution to Eq. \eqref{eq:FOM}, starting from an isotropic initial fiber orientation $\mathbf{A}(0) = \mathbf{I} / 3$, parameterized by the uncertainties $\omega = (T, L_f)$.
Here, $\mathbf{I} \in \R^{3 \times 3}$ denotes the identity matrix and $t_\text{end} > 0$ the end time, or time horizon, at which the fractional anisotropy is evaluated.
For instance, in an injection molding simulation, this could correspond to the moment when the part is fully filled.
Both response surfaces, one obtained using the FTE and the other using the iARD model, are shown in Figure \ref{fig:response_surface}.
For this, the analytical solutions to the corresponding FOMs from Appendix \ref{apx:analytical_sols} were utilized.
The curves show a smooth dependence of the fractional anisotropy $F$ on temperature and fiber length.
This behavior is anticipated, since the geometry, flow, and their corresponding functional dependencies on these factors exhibit continuous and differentiable behavior.
Specifically, at lower temperatures $T$, higher viscosity $\eta(T)$ aligns fibers more strongly in $(x_0, y_0)$, leading to increased anisotropy.
Conversely, shorter fibers are more prone to rotation, resulting in decreased anisotropy in their orientation.
Both curves appear similar, with the iARD model's response surface shifted upwards in parallel, indicating a higher anisotropy in general.
In qualitative terms, the uncertainties considered here appear to have the same influence on the QoI.
For the remainder of this study, curves generated using the FTE are depicted in blue, while those produced with the iARD model are shown in orange.

\subsubsection{Numerical solver}
Although analytical solutions for the considered FOMs are available, numerical approximations are employed to demonstrate the accuracy of the multilevel method.
For each $\omega = (T, L_f) \in \Omega$, we solve Eq. \eqref{eq:FOM} with initial isotropic $\mathbf{A}(0) = \mathbf{I} / 3$ using the forward Euler method as follows.
Given a time horizon $t_\text{end} > 0$ and a number of time steps $n \in \N$, define the step-size $\Delta t = t_\text{end} / n$.
Starting with an initial value $\mathbf{A}_0 := \mathbf{A}(0)$, we compute
\begin{equation}
    \mathbf{A}_{i+1} = \mathbf{A}_i + \Delta t \cdot M(\mathbf{A}_i;\omega)
\end{equation}
at time steps $t_i = i \Delta t$ for $i \in \{0, \ldots, n\}$, where $\omega \in \Omega$ denotes some realization of random uncertainties.
The corresponding approximate response surfaces are then given by
\begin{equation}
Q_n(\omega) := F(\mathbf{A}_n),
\end{equation}
which represent the functions from Section \ref{sec:ml_weighted_least_squares}.
Of course, higher-order time-stepping schemes can also be utilized.
To maximize flexibility in polynomial construction, we choose tensor product index sets, Eq. \eqref{eq:tensor_product}.

\subsubsection{Reference values for exact statistics}\label{sec:ref}
To compute the errors needed for verification of the bounds from Section \ref{sec:ml_l2_error}, we utilize the analytical solutions from Appendix \ref{apx:analytical_sols}.
The eigenvalues required for evaluating the fractional anisotropy, Eq. \eqref{eq:fa}, are computed numerically.
Our experiments demonstrate that the error introduced by these computations is negligible compared to the error inherent in solving the FOM itself.
As the QoI statistics are not analytically available, they must be estimated.
For this purpose, we utilize a \mbox{quasi-Monte Carlo} estimation on the Sobol sequence 
\begin{equation}\label{eq:sobol}
    \omega_S := \{\omega_i\}_{i=1}^N \subset \Omega
\end{equation}
with $N = 10^6$ samples \cite{Sobol1967}.
This sample set is employed to estimate the exact expected value, standard deviation, CDF, and $p$-quantile, as well as the surrogate's CDF and $p$-quantile.
The surrogate's expectation and standard deviation are computed exactly using the post-processing formulas from Section \ref{sec:comp_stats}.
Note that the derived bounds apply to the exact surrogate statistics, while we compute Monte Carlo approximations to some of these, such as for the CDF and the $p$-quantile.
However, our experiments indicate that the approximation based on the Sobol sample $\omega_S$ with $N = 10^6$ is sufficiently accurate to observe the predicted bounds.

\subsection{Results and discussion for fixed material point}
\begin{figure*}[h!]
  \centering
   \includegraphics[width=\linewidth]{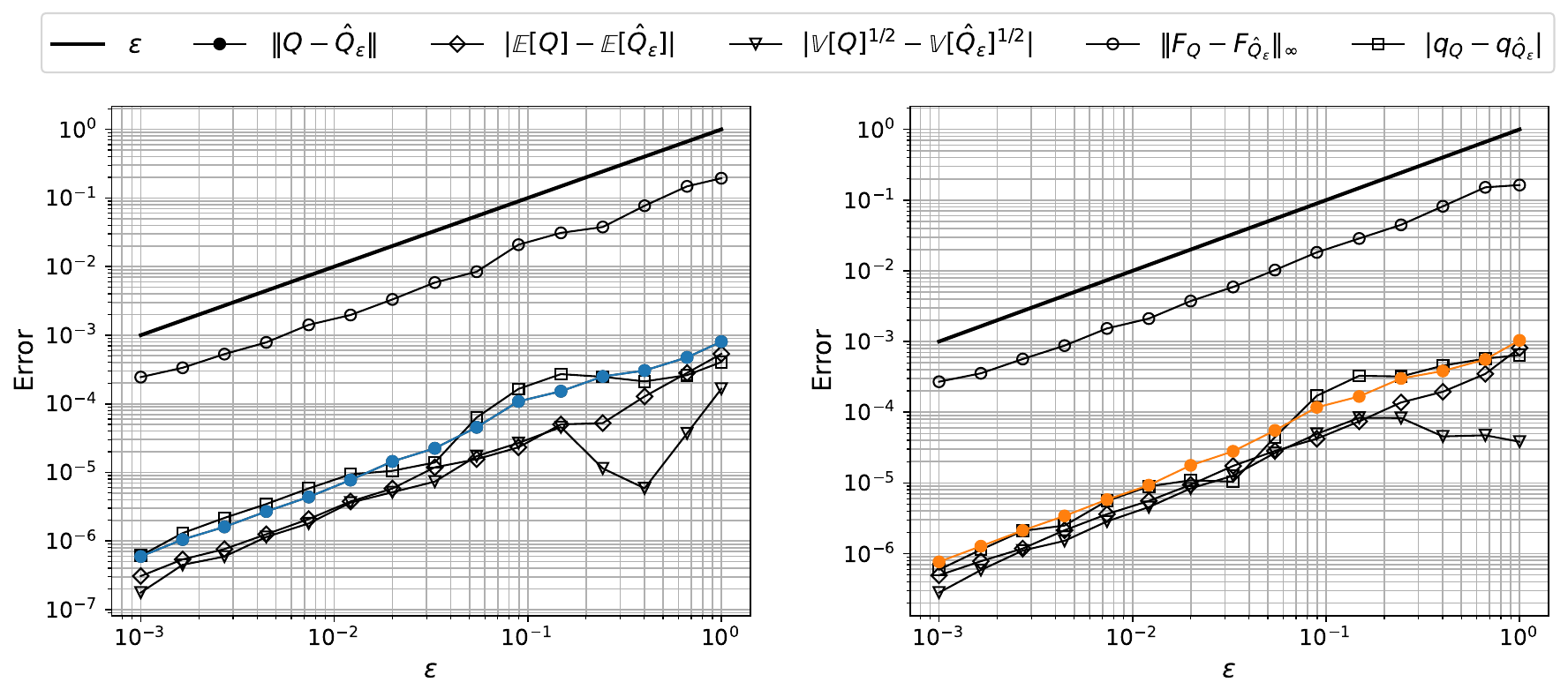}
    \caption{Input tolerance $\epsilon$ vs. the resulting error $\| Q - \hat{Q}_\epsilon\|$ of the polynomial approximation of fractional anisotropy $Q = F(\mathbf{A})$. Also shown are the errors for the expected value $\E[Q]$, standard deviation $\V[Q]^{1/2}$, CDF $F_Q$, and the $p$-quantile $q_Q$. The left plot presents results based on the FTE, while the right plot corresponds to the iARD model.}
    \label{fig:error_bounds}
\end{figure*}
\begin{figure}[h!]
    \centering
    \includegraphics[width=.5\linewidth]{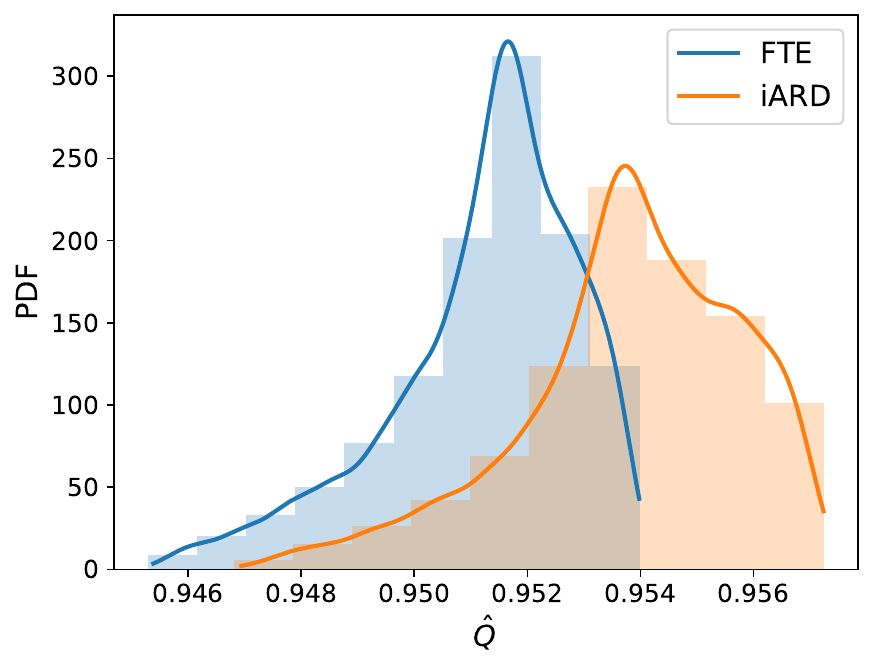}
    \caption{Shown are the PDFs and histograms of the QoI, fractional anisotropy, for the FTE and the iARD model. They were computed using a KDE of evaluations of their polynomial surrogates $\hat{Q}_\epsilon$ with $\epsilon = 10^{-3}$ on the Sobol sample $\omega_S$, Eq. \eqref{eq:sobol}.}
    \label{fig:pdfs}
\end{figure}

To numerically observe the predicted order of the bounds, we first select fifteen tolerance values, $\epsilon$, logarithmically distributed between $10^{-3}$ and $10^0$.
For each tolerance value $\epsilon$, we construct a polynomial surrogate $\hat{Q}_\epsilon$ using the multilevel least squares estimation, Eq. \eqref{eq:ml_lsq}.
On an Intel Core i5-1335U, the time for constructing the most accurate surrogate with $\epsilon = 10^{-3}$ was 42 s for the FTE and 61 s for the iARD model.
The asymptotic costs of computing $\hat{Q}_\epsilon$ are presented and illustrated in Appendix \ref{apx:complexity}, Figure \ref{fig:complexity_plot} and \ref{fig:complexity_plot_iard}.
As soon as a surrogate is available, the estimation of the statistics is achieved with minimal computational effort, consisting of polynomial evaluation on a Monte Carlo sample, which is negligible compared to the costs of constructing the surrogate itself.
For each surrogate, the error for the expectation
\begin{equation}
    |\E[Q] - \E[\hat{Q}_\epsilon]|
\end{equation}
is computed using the reference value from the previous Section \ref{sec:ref} and the post-processing formula from Section \ref{sec:comp_stats}.
This is also done for the error in standard deviation $\V[Q]^{1/2}$ and $p$-quantiles $q_Q$.
The uniform error for the CDFs,
\begin{equation}
    \|F_Q - F_{\hat{Q}_\epsilon}\|_\infty,
\end{equation}
is approximated by the maximum difference evaluated at the points of the Sobol sample $\omega_S$, Eq. \eqref{eq:sobol}.
For the FTE, the result is depicted in the left plot of Figure \ref{fig:error_bounds}, which shows that
\begin{equation}
    \| Q - \hat{Q}_\epsilon\| \leq \epsilon,
\end{equation}
represented by the blue curve, for all trained polynomial surrogates.
It also verifies the theoretical bounds from Section \ref{sec:ml_l2_error}. The right plot of the same Figure \ref{fig:error_bounds} shows the same curves for the iARD model.
All error curves for the statistics, represented by the curves with non-solid markers, are essentially parallel to the solid $\epsilon$-curve.
The predicted order of $2/3$, representing the slope of the error curve, in Ineq. \eqref{eq:cdf_err} for the CDF (and thus, quantile) error is not evident here, likely because the underlying problem is well-posed in the sense of smooth dependence of the uncertainties on the QoI.
Note that this does not contradict the theory, as Ineq. \eqref{eq:cdf_err} is only an upper bound.
The upward shift of the CDF curve, compared to the other statistics, can also be explained by this bound.
The constant includes the maximum value $\|f_Q\|_\infty$ of the associated density, which is approximately $500$ in this case and can be read off in Figure \ref{fig:pdfs}.
The factor $f_Q(q)^{-1}$ in Eq. \eqref{eq:quantile_bound} for the quantile bound then scales the quantile error curve down again.

Table \ref{tbl:results} presents the reference values for the statistical quantities for the QoI computed using the FTE and the iARD model.
The iARD model uses the same fiber interaction coefficient $C_I = 0.01$ as FTE and anisotropy coefficient $C_M = 0.2$.
Note that these values have been chosen arbitrarily.
The table compares the expected value $\E[Q]$, standard deviation $\V[Q]^{1/2}$, CDF at an arbitrarily chosen point $F_Q(0.953)$, and the $p$-quantile $q_Q$ with $p = 0.99$ for both models, highlighting the relative errors in each case.
The relative errors were calculated using the polynomial surrogate $\hat{Q}_\epsilon$ with $\epsilon = 10^{-3}$.
In general, the results show low relative errors in the approximations, which underscores the accuracy of the surrogate.
The fractional anisotropy indicates a high alignment of $0.9512$ for the FTE model and $0.9538$ for the iARD model in the expectation.
Under uncertainty in the material properties of temperature and fiber length, the standard deviation for the iARD model is $1.9665 \cdot 10^{-3}$, which is slightly higher than the FTE standard deviation of $1.6794 \cdot 10^{-3}$.
This indicates greater variability in the fractional anisotropy predicted by the iARD model.
\begin{table*}[h!]
\centering
\begin{tabular}{llllllllll}
\toprule
& \multicolumn{2}{c}{$\E[Q]$} & \multicolumn{2}{c}{$\V[Q]^{1/2}$} & \multicolumn{2}{c}{$F_Q(0.953)$} & \multicolumn{2}{c}{$q_Q$} \\
\cmidrule(lr){2-3} \cmidrule(lr){4-5} \cmidrule(lr){6-7} \cmidrule(lr){8-9}
& ref. value & rel. error & ref. value & rel. error & ref. value & rel. error & ref. value & rel. error \\
\midrule
\textbf{FTE} & 0.9512 & $3.237 \cdot 10^{-7}$ & $1.6794 \cdot 10^{-3}$ & $1.053 \cdot 10^{-4}$ & 0.8723 & $5.732 \cdot 10^{-5}$ & 0.9538 & $6.405 \cdot 10^{-7}$ \\
\textbf{iARD} & 0.9538 & $5.177 \cdot 10^{-7}$ & $1.9665 \cdot 10^{-3}$ & $1.429 \cdot 10^{-4}$ & 0.2821 & $3.190 \cdot 10^{-4}$ & 0.9570 & $6.182 \cdot 10^{-7}$ \\
\bottomrule
\end{tabular}
\caption{Comparison of expected value $\E[Q]$, standard deviation $\V[Q]^{1/2}$, CDF $F_Q$ at the arbitrarily chosen point $0.953$, and the $p$-quantile $q_Q$ with $p=0.99$ for the fractional anisotropy $Q = F(\mathbf{A})$ using the FTE and iARD model, including their reference values and relative errors. The relative errors were computed using the polynomial surrogate $\hat{Q}_\epsilon$ with $\epsilon = 10^{-3}$.}\label{tbl:results}
\end{table*}
This can also be observed in Figure \ref{fig:pdfs}, which depicts the probability density function (PDF) and histogram for both models, computed using a kernel density estimation (KDE) \cite{Rosenblatt1956} of the evaluation of their surrogates on the Sobol sample $\omega_S$, Eq. \eqref{eq:sobol}.
The iARD model's density function exhibits a broader support, indicating a higher variance.
Similar to what can be observed in Figure \ref{fig:response_surface}, the iARD model predicts a higher fractional anisotropy.

\subsection{Entire flow channel}\label{sec:exp_2}
\begin{figure*}[h!]
\centering
\begin{subfigure}{.5\textwidth}
  \centering
  \includegraphics[width=\linewidth]{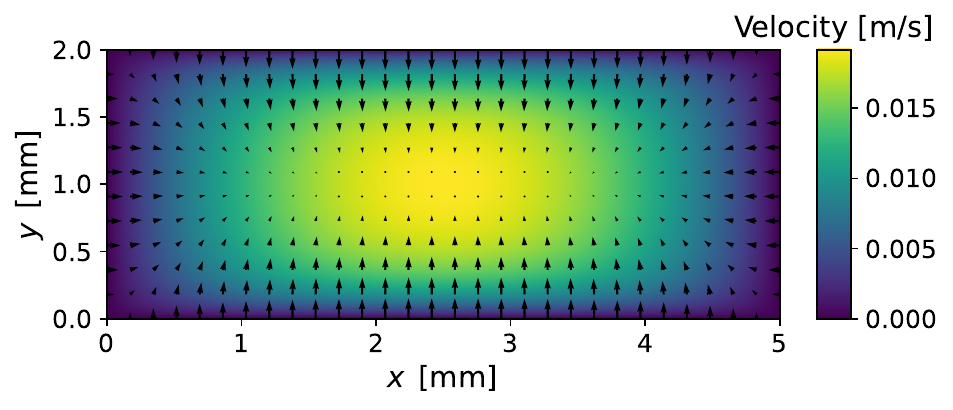}
  \caption{Hagen-Poiseuille flow}
  \label{fig:hp_flow}
\end{subfigure}%
\begin{subfigure}{.5\textwidth}
  \centering
  \includegraphics[width=\linewidth]{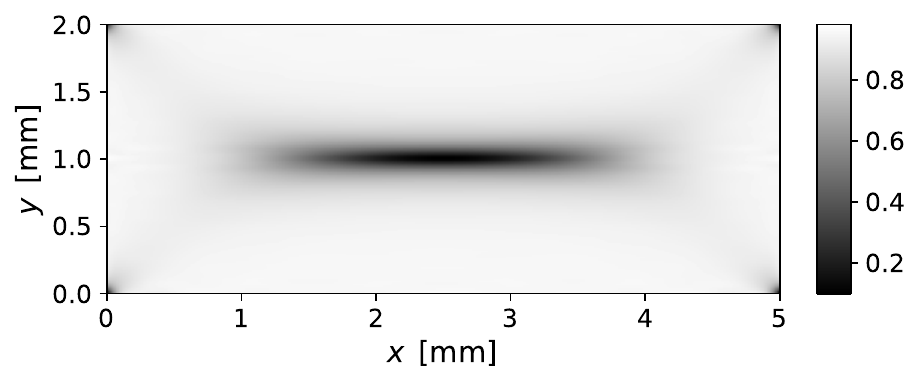}
  \caption{Expected value $\E[\hat{Q}_{(x, y)}]$}
  \label{fig:means}
\end{subfigure}
\begin{subfigure}{.5\textwidth}
  \centering
  \includegraphics[width=\linewidth]{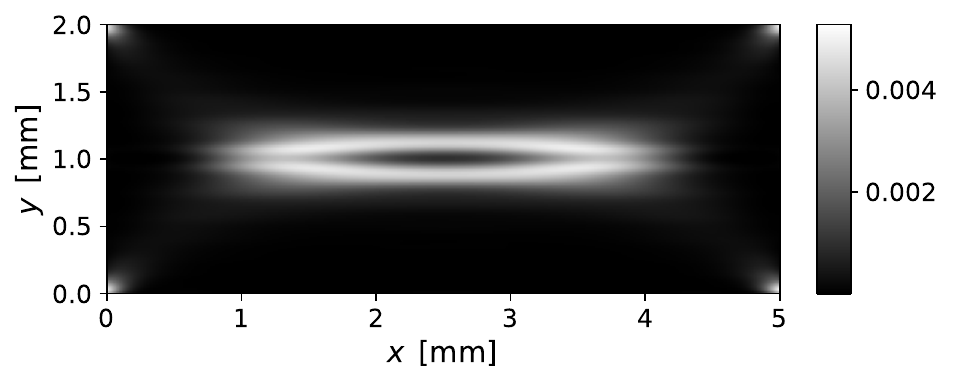}
  \caption{Standard deviation $\V[\hat{Q}_{(x, y)}]^{1/2}$}
  \label{fig:stds}
\end{subfigure}%
\begin{subfigure}{.5\textwidth}
  \centering
  \includegraphics[width=\linewidth]{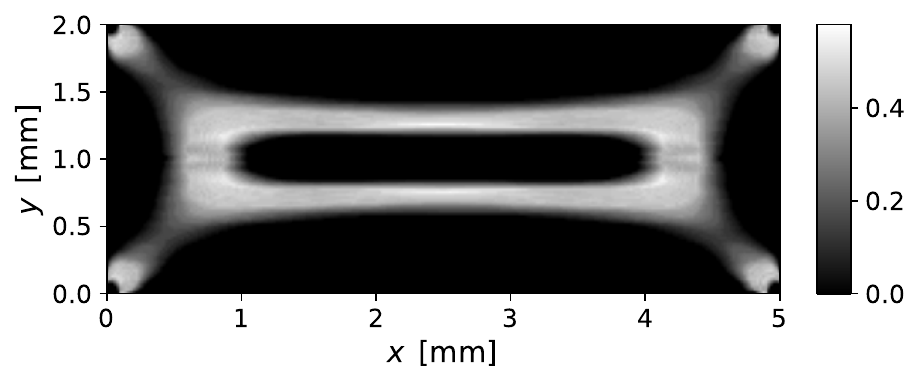}
  \caption{Probability $\mathbb{P}(0.85 \leq \hat{Q}_{(x, y)} \leq 0.9)$}
  \label{fig:probs}
\end{subfigure}
\caption{Hagen-Poiseuille flow (\ref{fig:hp_flow}) and surrogate statistics of the fractional anisotropy $\hat{Q}_{(x, y)} = F(A)$, according to Eq. \eqref{eq:fa}, under uncertain temperature and fiber length, across the channel (\ref{fig:means} - \ref{fig:probs}). All computations were performed using the FTE, Eq. \eqref{eq:fte}, as the fiber orientation model.}
\end{figure*}

In this second part of the experiment, we extend the approach to the entire flow channel as described in Section \ref{sec:global_surrogates}.
Using the notation defined there, let \mbox{$\Omega = \Omega_T \times \Omega_{L_f}$} still denote the space of random temperature and fiber length, while $\mathcal{X} = [0, w] \times [0, h]$ denotes the channel domain.
To demonstrate the effectiveness of this augmented approach, we consider a grid of $200 \times 200$ equidistant points in the channel $\mathcal{X} = [0, w] \times [0, h]$.
All computations utilize the FTE.
The only parameter altered from the previous part is the reduction of the time horizon $t_\text{end}$ from $300$ to $200$ seconds, which was implemented to enhance the visualization of variance differences across various points in the channel.
As the time horizon is shorter, the steady state of an FOM has not been reached yet to the same extent, and uncertainties are reflected more by larger variability of the QoI.
It is important to note that, in general, the polynomial approximation of functions remains feasible as long as the number of random parameters, is moderate, avoiding the curse of dimensionality \cite{Bellman1966}.
In this experiment, we have $d + p = 4$ variables, which can be considered small.
For the construction of $\hat{Q}$, the adaptive version of the multilevel least squares method with $200$ number of steps has been utilized.
This results in $\|Q - \hat{Q}\| \approx 3.503 \cdot 10^{-3}$, numerically verified through a quasi-Monte Carlo method with Sobol sequence of size $N=10^{4}$.
For each grid point $(x, y) \in \mathcal{X}$, the partial evaluation 
\begin{equation}
    \hat{Q}_{(x, y)} := \hat{Q}\left(\cdot, (x, y)\right)
\end{equation}
was computed, and the corresponding expectations are shown in Figure \ref{fig:means}.
The expectations and variances are computed exactly by post-processing the Legendre coefficients, as described in Section \ref{sec:comp_stats}.

\subsection{Results and discussion for the entire flow channel}\label{sec:full_channel_learning}
It can be observed that the expected fractional anisotropy spans almost the entire range from $0$ to $1$ across the channel.
This is primarily influenced by the Hagen-Poiseuille flow and its associated velocity gradient, Eq. \eqref{eq:velocity_grad}.
For example, the horizontal middle strip of the channel exhibits low anisotropy due to the small values in the velocity gradient, which do not significantly reorient the fibers (see Figure \ref{fig:hp_flow}).
This effect is also noticeable in the corners of the channel.
Figure \ref{fig:stds} depicts the standard deviation within the channel.
The QoI, fractional anisotropy, is not significantly impacted by uncertain temperature or fiber length in areas with sufficiently high velocity gradient, resulting in low standard deviation.
Conversely, in regions where the gradient is almost zero, fiber reorientation is minimal, so uncertainties in these parameters also have little effect on the anisotropy, as observed in the slim horizontal middle strip of the channel.
Of particular interest is the ring surrounding this middle strip, which exhibits the highest standard deviation.
In this area, the material properties, temperature and fiber length, have the most pronounced influence on the QoIs variability.
This region likely represents a zone where the underlying velocity gradient is neither too low nor too high, allowing the effects of uncertainties in material properties to become more apparent.
Finally, Figure \ref{fig:probs} depicts the CDF evaluation at arbitrarily chosen points.
Specifically, we examine the probability that the fractional anisotropy lies between $0.85$ and $0.9$, which can be computed by
\begin{equation}
    \mathbb{P}(0.85 \leq \hat{Q}_{(x, y)} \leq 0.9) = F_{\hat{Q}_{(x, y)}}(0.9) - F_{\hat{Q}_{(x, y)}}(0.85),
\end{equation}
where $F_{\hat{Q}_{(x, y)}}$ is the CDF of the surrogates partial evaluation.
Gradual oscillations can be observed, which likely result from the polynomial approximation of the response surface.
Similar to the standard deviation, a ring or contour line is present, within which the considered values of the QoI can be found, supported by probabilistic information.

\section{Conclusion}\label{sec:conclusion}
This study focuses on modeling fiber orientation in composite molding processes using novel error bounds for statistics of a quantity of interest (QoI) computed via multilevel polynomial surrogates.
These bounds are based on the $L^2$ error of the polynomial approximation.
We utilize polynomials with controllable error through multilevel techniques, while also being significantly more efficient than traditional Monte Carlo estimators.
Numerical experiments employing the Cross-WLF viscosity model, Hagen-Poiseuille flow, the Folgar-Tucker equation (FTE), and the improved anisotropic rotary diffusion (iARD) model verify the proposed theory.
Our findings show that uncertainties in material properties, such as temperature and fiber length, have significant impact on the fractional anisotropy of fiber orientation.
We observe that both the FTE and the iARD model produce similar results at first glance for deterministic simulations.
However, with polynomial surrogates, it is possible to reveal subtle yet statistically significant differences between these fiber orientation models, which highlights another aspect of their usefulness.
In safety-critical applications, even small deviations can impact reliability and performance of fiber-reinforced polymers (FRPs), such that accurately quantifying these statistical risk factors is necessary for informed decision-making.
It is important to note that our method enables meaningful model comparisons because it provides error bounds for our estimates.
For instance, when comparing the $p$-quantile of fractional anisotropy between two fiber orientation models, the comparison is only reliable if the errors of the estimates are of the same order of magnitude.
Our derived error bounds guarantee this consistency.

In addition to investigating the fiber orientation distribution in a single material point, we extended the approach to learn the spatial distribution of the fractional anisotropy across the entire rectangular channel domain.
This enables the investigation of how uncertain material and process parameters influence fiber orientation throughout the entire part’s domain.
By doing so, it becomes possible to precisely identify regions within a FRP part that exhibit high uncertainty in fiber orientation, such as areas with high variance.
Such variations in fiber orientation can lead to safety-critical issues due to inconsistencies in the mechanical properties of the produced part.
Another potential application is the computation of failure probabilities, such as the likelihood of an FRP part failing due to insufficient fiber alignment.
Our extension enables the evaluation of these failure probabilities across the entire parts domain.
Again, the reliability of the estimated failure probabilities is given by the provided error bounds, which in turn enables a comparison (of the failure probabilities) of different fiber orientation models across the FRP part.
To support the geometry-augmented approach, theoretical bounds for the $L^2$ error control of statistics over the entire domain through partial evaluations are presented.
These considerations indicate that while the derivation of a global, geometry-wide mapping of the QoI to pointwise evaluations is limited from a theoretical point of view, it is still effective for qualitative investigations of fiber orientation.
One limitation of the presented approach is its reliance on finite-dimensional uncertainties, instead of, for example, time-varying random fields.
Additionally, the numerical experiments are conducted on relatively simple geometries with smooth response surfaces, where polynomial surrogates perform well.
Their performance may drop in more complicated settings with non-smooth responses.

Despite these limitations, the derived error bounds are not inherently tied to polynomial surrogates and could extend to alternative methods, such as neural networks.
Nonetheless, the multilevel polynomial least squares approach retains a distinct advantage through its a-priori $L^2$ error bounds, providing a strong theoretical foundation.
Overall, this research contributes to the understanding and application of polynomial surrogates and provides a robust framework for uncertainty quantification, error control and model comparison in the context of fiber orientation.

\clearpage
\nomenclature[M, 01]{$\mathbf{I}$}{Identity matrix}
\nomenclature[M, 02]{$\mu$}{Probability measure}
\nomenclature[M, 03]{$\Omega$}{Uncertainty domain}
\nomenclature[M, 04]{$\omega$}{Realization of uncertain parameters}
\nomenclature[M, 05]{$u$}{Solution to computational model}
\nomenclature[M, 06]{$q$}{Quantity of interest}
\nomenclature[M, 07]{$Q$}{Response surface}
\nomenclature[M, 08]{$Q_n$}{Approximative response surface}
\nomenclature[M, 09]{$\hat{Q}$}{Polynomial approximation to response surface}
\nomenclature[M, 10]{$\lambda, \Lambda$}{Multi-index / Multi-index set}
\nomenclature[M, 11]{$P_\lambda$}{Polynomial corresponding to multi-index $\lambda$}
\nomenclature[M, 12]{$L^2_\mu(\Omega)$}{Space of functions on $\Omega$ with finite $L^2_\mu$ norm}
\nomenclature[M, 13]{$\Pi^N, \Pi$}{Discrete / continuous least squares operator}
\nomenclature[M, 14]{$d\nu/d\mu$}{Radon–Nikodym derivative}
\nomenclature[M, 15]{$\epsilon$}{Error tolerance}
\nomenclature[M, 16]{$\E$}{Expected value}
\nomenclature[M, 17]{$\V$}{Variance}
\nomenclature[M, 18]{$\mathbb{P}$}{Probability}
\nomenclature[M, 19]{$f_X$}{Probability density function of random variable $X$}
\nomenclature[M, 20]{$F_X$}{Cumulative distribution function of random variable $X$}
\nomenclature[M, 21]{$q_X$}{$p$-Quantile of random variable $X$}

\nomenclature[F, 01]{$v$}{Velocity}
\nomenclature[F, 02]{$P$}{Pressure gradient}
\nomenclature[F, 03]{$\eta$}{Viscosity}
\nomenclature[F, 04]{$T$}{Temperature}
\nomenclature[F, 05]{$\dot{\gamma}$}{Shear rate}
\nomenclature[F, 06]{$\mathcal{S}_+$}{Space of symmetric and positive semi-definite matrices with trace one}
\nomenclature[F, 07]{$\mathbf{A}, \mathbb{A}$}{Second / Fourth order fiber orientation tensor}
\nomenclature[F, 08]{$\mathbf{L}$}{Velocity gradient}
\nomenclature[F, 09]{$\mathbf{D}, \mathbf{W}$}{Rate-of-deformation / vorticity tensor}
\nomenclature[F, 10]{$\xi$}{Particle shape factor}
\nomenclature[F, 11]{$C_I$}{Fiber interaction coefficient}
\nomenclature[F, 12]{$C_M$}{Anisotropy coefficient}
\nomenclature[F, 13]{$\mathbf{C}$}{Rotary diffusion tensor}
\nomenclature[F, 14]{$F$}{Fractional anisotropy}
\nomenclature[F, 15]{$L_f, D_f$}{Fiber length / diameter}
\nomenclature[F, 16]{$M$}{General fiber orientation model}
\nomenclature[F, 17]{$\Delta t$}{Step size}
\nomenclature[F, 18]{$t_\text{end}$}{Time horizon}
\nomenclature[F, 19]{$\mathcal{X}$}{Parts domain}

\renewcommand{\nomname}{List of Symbols}
\begin{small}
\printnomenclature[0.5in]
\end{small}

\clearpage
\appendix
\section{Analytical solution to fiber orientation models}\label{apx:analytical_sols}
We utilize the analytical solution derived by \citet{Winters2022} and present the exact formulas needed to perform the numerical experiments in this work.
Let $\mathbf{e} = \mathbf{e}_1 + \mathbf{e}_5 + \mathbf{e}_9$, where $\mathbf{e}_i \in \R^9$ denotes the unit vector with a 1 in the $i$-th position and 0 elsewhere.
Further, let $\texttt{vec}: \R^{3 \times 3} \rightarrow \R^9$ be the mapping for the column-wise vectorization of a matrix into a vector, and $\texttt{mat}: \R^9 \rightarrow \R^{3 \times 3}$ the reverse operation.
The following analytical solutions are valid for a hybrid closure \cite{Advani1990}, defined as
\begin{equation}
    \mathbb{A}^H = (1-f)\mathbb{A}^L + f \mathbb{A}^Q,
\end{equation}
where $f \in [0, 1]$, and $\mathbb{A}^Q_{ijkl} = \mathbf{A}_{ij} \mathbf{A}_{kl}$ represents the quadratic closure \cite{Doi1981}, and
\begin{equation}
    \mathbb{A}^L_{ijkl} = -\frac{1}{35} \left( \delta_{ij} \delta_{kl} \!+\! \delta_{ik} \delta_{jl} \!+\! \delta_{il} \delta_{jk} \right) + \frac{1}{7} \left( \mathbf{A}_{ij} \delta_{kl} \!+\! \mathbf{A}_{ik} \delta_{jl} \!+\! \mathbf{A}_{il} \delta_{jk} \!+\! \mathbf{A}_{jl} \delta_{ik} \!+\! \mathbf{A}_{jk} \delta_{il} \!+\! \mathbf{A}_{kl} \delta_{ij} \right)
\end{equation}
denotes the linear closure \cite{Hand1962}.
Note that $f$ must not depend on $\mathbf{A}$, which excludes the classical hybrid closure with $f = 1 - 27 \, \text{det}(\mathbf{A})$.
Define the variable $\mathbf{L_\xi} := \mathbf{W} + \xi \mathbf{D}$, the linear mapping $\mathcal{L}: \R^{3 \times 3} \rightarrow \R^{9 \times 9}$ with
\begin{equation}
    \mathcal{L}(\mathbf{X}) = \frac{1}{7} \text{tr}(\mathbf{X}) \, \mathbf{I} + \frac{2}{7}(\mathbf{I} \otimes \mathbf{X} + \mathbf{X} \otimes \mathbf{I}) + \frac{1}{7} \mathbf{e} \, \texttt{vec}(\mathbf{X})^\top - \frac{2}{35} \texttt{vec}(\mathbf{X}) \,\mathbf{e}^\top - \frac{1}{35} \text{tr}(\mathbf{X}) \,\mathbf{e}\mathbf{e}^\top,
\end{equation}
as well as the trace normalizing function $\text{tr}^*(\mathbf{X}) = \mathbf{X} / \text{tr}(\mathbf{X})$ for a matrix $\mathbf{X} \in \mathbb{R}^{n \times n}$.
Using the notation introduced above, the solution to the Folgar-Tucker equation, Eq. \eqref{eq:fte}, with initial orientation tensor $\mathbf{A}_0 \in \R^{3 \times 3}$ is given by
\begin{equation}
    \mathbf{A}(t) = \text{tr}^* \left( \texttt{mat}\left(e^{t\mathbf{M}} \, \texttt{vec}\left(\mathbf{A}_0\right)\right)\right)
\end{equation}
with
\begin{equation}
    \mathbf{M} = \mathbf{I} \otimes \mathbf{L_\xi} + \mathbf{L_\xi} \otimes \mathbf{I} +  2C_I \dot{\gamma} \left(\mathbf{e}\mathbf{e}^\top - 3 \mathbf{I}\right) - 2 \xi (1 - f) \mathcal{L}(\mathbf{D}).
\end{equation}
Similarly, defining $\mathbf{E} := \xi \mathbf{D} - 5 \dot{\gamma}\mathbf{C}$ as suggested by \citet{Favaloro2019}, the solution to the anisotropic rotary diffusion model, Eq. \eqref{eq:ard}, with initial orientation tensor $\mathbf{A}_0 \in \R^{3 \times 3}$ is given by
\begin{equation}
    \mathbf{A}(t) = \text{tr}^* \left( \texttt{mat}\left(e^{t\mathbf{M}} \, \texttt{vec}\left(\mathbf{A}_0\right)\right)\right)
\end{equation}
with
\begin{equation}
    \mathbf{M} = \mathbf{I} \otimes \mathbf{L_\xi} + \mathbf{L_\xi} \otimes \mathbf{I} + 2 \dot{\gamma} \left( \texttt{vec}(\mathbf{C}) \, \mathbf{e}^\top - \text{tr}(\mathbf{C}) \, \mathbf{I}\right) - 5 \dot{\gamma} (\mathbf{I} \otimes \mathbf{C} + \mathbf{C} \otimes \mathbf{I}) - 2(1 - f) \mathcal{L}(\mathbf{E}).
\end{equation}

\section{Error bounds for statistical quantities}\label{apx:error_bounds}
In the following sections, we provide a detailed explanation on the derivation of error bounds for several statistics that are approximated using a surrogate $\hat{Q}$.
The aim in each case is to ensure that the upper bound includes the $L^2$ approximation error $\|Q - \hat{Q}\|$ of the surrogate.

\subsection{Standard deviation}\label{apx:std}
Using the reverse triangle inequality it follows for the difference of the standard deviations
\begin{align}\label{eq:std_bound}
\begin{split}
    \left|\V[Q]^{1/2} - \V[\hat{Q}]^{1/2}\right|^2
    &= \left|\|Q - \E[Q]\| - \|\hat{Q} - \E[\hat{Q}]\|\right|^2 \\
    &\leq \|(Q - \hat{Q}) - (\E[Q] - \E[\hat{Q}])\|^2 \\
    &= \|Q - \hat{Q}\|^2 + \|\E[Q] - \E[\hat{Q}]\|^2 - 2 \langle Q - \hat{Q}, \,\E[Q] - \E[\hat{Q}]\rangle \\
    &= \|Q - \hat{Q}\|^2 - |\E[Q] - \E[\hat{Q}]|^2 \\
    &\leq \|Q - \hat{Q}\|^2,
\end{split}
\end{align}
where $\langle f, g\rangle = \int_\Omega f(\omega) g(\omega) \, \mu(d \omega)$ denotes the $L^2_\mu$ inner product for elements $f, g \in L^2_\mu(\Omega)$.

\subsection{Cumulative distribution function}\label{apx:cdf}
In addition to the $L^2$ norm considered so far, we define, for $p \in (0, \infty)$, the $L^p$ norm as
    \begin{equation}
    \|f\|_{L^p_\mu(\Omega)} = \left(\int_\Omega |f(\omega)|^p \,\mu(d\omega)\right)^\frac{1}{p}
\end{equation}
and abbreviate $\|\cdot\|_{L^p} := \|\cdot\|_{L^p_\mu(\Omega)}$.
\begin{lemma}\label{lemma:avikainen}
Let $X$ and $Y$ be random variables on the same probability space.
Then, if $X$ has a bounded density $f_X$, it holds for every $0 < p < \infty$ that
\begin{equation}
    \sup_{x \in \R} \E \bigl[|\mathbbm{1}_{\{X \leq x\}} - \mathbbm{1}_{\{Y \leq x\}}| \bigr] \leq 3 \left(\| f_X \|_\infty \E \bigl[ |X-Y|^p\bigr]^\frac{1}{p}\right)^{\frac{p}{p+1}}.
\end{equation}
\end{lemma}
\newproof{pf}{Proof}
\begin{pf}
    See Lemma 3.4 in \cite{Avikainen2009}.
    \qed
\end{pf}
Denote by $F_X := \mathbb{P}(X \leq \cdot)$ the cumulative distribution function (CDF) of a random variable $X$.
The $L^2$ convergence of the polynomial surrogate $\hat{Q}$ towards $Q$ implies convergence in distribution, which means that $F_{\hat{Q}}(x) \rightarrow F_Q(x)$ for all $x \in \R$ at which $F_Q$ is continuous.
By applying the preceding lemma, a more quantitative result can be obtained.
For the error between the CDFs of $Q$ and $\hat{Q}$, it holds that
\begin{equation}
    \left| F_Q(x) - F_{\hat{Q}}(x) \right| = \left| \E[\mathbbm{1}_{\{Q \leq x\}}] - \E[\mathbbm{1}_{\{\hat{Q} \leq x\}}]\right| \leq \E\left[\left| \mathbbm{1}_{\{Q \leq x\}} - \mathbbm{1}_{\{\hat{Q} \leq x\}} \right|\right]
\end{equation}
for all $x \in \R$.
Taking the supremum on both sides and applying Lemma \ref{lemma:avikainen} with $X=Q$ and $Y=\hat{Q}$, as well as $p=2$, yields
\begin{equation}\label{eq:cdf_bound}
    \| F_Q - F_{\hat{Q}}\|_{\infty} \leq 3 \left(\| f_Q \|_{\infty} \| Q-\hat{Q}\|\right)^{\frac{2}{3}}.
\end{equation}

\subsection{Quantile}
For a probability $p \in (0, 1)$, denote by
\begin{equation}
    q_X := \inf \{q \in \R: F_X(q) \geq p\}
\end{equation}
the $p$-quantile of a random variable $X$.
\begin{lemma}\label{lemma:quantile}
    Let $X$ and $Y$ be random variables with CDFs $F_X, F_Y$ and $p$-quantiles $q_X, q_Y$, for $p \in (0, 1)$, respectively.
    Then, if $X$ has a density with $f_X > 0$ on $\emph{conv}(q_X, q_Y)$,
    it holds that
    \begin{equation}
        |q_X - q_Y| \leq f_X(q)^{-1} \|F_X - F_Y\|_{\infty}
    \end{equation}
    for some intermediary $q \in \emph{conv}(q_X, q_Y)$.
\end{lemma}
\begin{pf}
    According to the mean value theorem, there exists some $q \in \text{conv}(q_X, q_Y)$ such that
    \begin{equation}
        \frac{|F_X(q_X) - F_X(q_Y)|}{|q_X - q_Y|} = f_X(q).
    \end{equation}
    Rearranging, substituting $F_X(q_X) = F_Y(q_Y) = p$ by definition of the $p$-quantile, and taking the supremum yields the claim.
    \qed
\end{pf}
Again, by $L^2$ convergence (and thus convergence in distribution), one has $q_{\hat{Q}} \rightarrow q_Q$ if $F_Q$ is continuous in $q_Q$.
By Lemma \ref{lemma:quantile} and Ineq. \eqref{eq:cdf_bound}, we can bound the error between the $p$-quantiles of $Q$ and $\hat{Q}$ by
\begin{align}\label{eq:aux_quantile_bound}
\begin{split}
    |q_Q - q_{\hat{Q}}|
    &\leq f_Q(q)^{-1} \|F_Q - F_{\hat{Q}}\|_{\infty} \\
    &\leq 3 f_Q(q)^{-1} \left(\| f_Q \|_{\infty} \| Q-\hat{Q}\|\right)^{\frac{2}{3}}
\end{split}
\end{align}
for some $q \in \text{conv}(q_Q, q_{\hat{Q}})$.

\section{Extension of error bounds for statistical quantities}\label{apx:error_bounds_ext}

\subsection{Expected value}
Let $m(x) := \E[Q(\cdot, x)]$ and $\hat{m}(x) := \E[\hat{Q}(\cdot, x)]$.
As with Ineq. \eqref{eq:jensen}, we use Jensen's inequality to bound
\begin{align}
    \begin{split}
        \|m - \hat{m}\|^2_{L^2_{\mu_x}}
        &= \int_\mathcal{X} |m(x) - \hat{m}(x)|^2 \mu_x(dx) \\
        &= \int_\mathcal{X} \left|\int_\Omega Q(\omega, x) - \hat{Q}(\omega, x) \, \mu_\omega(d \omega)\right|^2 \mu_x(dx) \\
        &\leq \int_\mathcal{X} \int_\Omega \left | Q(\omega, x) - \hat{Q}(\omega, x) \right |^2 \mu_\omega \, (d \omega) \mu_x (dx) \\
        &= \|Q - \hat{Q}\|_{L^2_\mu}^2.
    \end{split}
\end{align}

\subsection{Standard deviation}
Let $s(x) := \V[Q(\cdot, x)]^{1/2}$ and $\hat{s}(x) := \V[\hat{Q}(\cdot, x)]^{1/2}$.
Then we use the error bound for standard deviations, Ineq. \eqref{eq:std_bound}, from Appendix \ref{apx:std} to obtain
\begin{align}
    \begin{split}
        \|s - \hat{s}\|^2_{L^2_{\mu_x}}
        &= \int_\mathcal{X} |s(x) - \hat{s}(x)|^2 \mu_x(dx) \\
        &= \int_\mathcal{X} \left|\V[Q(\cdot, x)]^{1/2} - \V[\hat{Q}(\cdot, x)]^{1/2}\right|^2 \mu_x(dx) \\
        &\leq \int_\mathcal{X} \int_\Omega \left | Q(\omega, x) - \hat{Q}(\omega, x) \right |^2 \mu_\omega \, (d \omega) \mu_x (dx) \\
        &= \|Q - \hat{Q}\|_{L^2_\mu}^2.
    \end{split}
\end{align}

\subsection{Cumulative distribution function}
Let $F_z(x) := \mathbb{P}(Q(\cdot, x) \leq z)$ and $\hat{F}_z(x) := \mathbb{P}(\hat{Q}(\cdot, x) \leq z)$ for $z \in \R$.
By Jensen's inequality it holds
\begin{align}
    \begin{split}
        \|F_z - \hat{F}_z\|^2_{L^2_{\mu_x}}
        &= \int_\mathcal{X}
        \left|\mathbb{P}(Q(\cdot, x) \leq z) - \mathbb{P}(\hat{Q}(\cdot, x) \leq z)\right|^2 \mu_x(dx) \\
        &= \int_\mathcal{X} \left|\E[\mathbbm{1}_{\{Q (\cdot, x)\leq z\}}] - \E[\mathbbm{1}_{\{\hat{Q} (\cdot, x)\leq z\}}]\right|^2 \mu_x(dx) \\
        &\leq \int_\mathcal{X} \int_\Omega \left| \mathbbm{1}_{\{Q (\omega, x)\leq z\}} - \mathbbm{1}_{\{\hat{Q} (\omega, x)\leq z\}}\right |^2 \mu_\omega \, (d \omega) \mu_x (dx) \\
        &= \|\mathbbm{1}_{\{Q \leq z\}} - \mathbbm{1}_{\{\hat{Q} \leq z\}}\|^2_{L^2_\mu}.
    \end{split}
\end{align}
Applying Theorem 2.4 (i) from \cite{Avikainen2009} and taking the supremum yields
\begin{equation}
    \sup_{z \in \R} \|F_z - \hat{F}_z\|^2_{L^2_{\mu_x}} \leq 27 \left(\| f_Q \|_{\infty} \| Q-\hat{Q}\|_{L^2_\mu} \right)^{\frac{2}{3}}.
\end{equation}

\subsection{Quantile}
Let $q(x) := q_{Q(\cdot, x)}$ and $\hat{q}(x) := q_{\hat{Q}(\cdot, x)}$ denote the $p$-quantile of $Q(\cdot, x)$ and $\hat{Q}(\cdot, x)$, respectively.
We assume that there exists $c > 0$ such that $f_{Q(\cdot, x)} \geq c$ on $\text{conv}\left(q(x), \hat{q}(x)\right)$ for all $x \in \mathcal{X}$.
Then, Ineq. \eqref{eq:aux_quantile_bound} states that
\begin{equation}
    |q(x) - \hat{q}(x)| \leq 3 c^{-1} \left(\| f_Q \|_{\infty} \| Q(\cdot, x)-\hat{Q}(\cdot, x)\|_{L^2_\mu}\right)^{\frac{2}{3}},
\end{equation}
for all $x \in \mathcal{X}$, where we have also used $\| f_{Q(\cdot, x)} \|_{\infty} \leq \| f_Q \|_{\infty}$.
We can conclude
\begin{align}
\begin{split}
    \|q-\hat{q}\|_{L^2_{\mu_x}}^2
    &= \int_\mathcal{X} |q(x) - \hat{q}(x)|^2 \mu_x(dx) \\
    &\leq 9c^{-2} \| f_Q \|_{\infty}^{\frac{4}{3}} \int_\mathcal{X} \left ( \|Q(\cdot, x) - \hat{Q}(\cdot, x)\|^2_{L^2_{\mu_\omega}}\right )^{\frac{2}{3}} \mu_x(dx) \\
    &\leq 9 c^{-2} \| f_Q \|_{\infty}^{\frac{4}{3}} \left( \|Q - \hat{Q}\|^2_{L^2_{\mu}}\right)^{\frac{2}{3}},
\end{split}
\end{align}
where in the last step we applied Jensen's inequality reversely to the concave function $|\cdot|^{2/3}$.
Thus, in total
\begin{equation}
    \|q-\hat{q}\|_{L^2_{\mu_x}} \leq 3 c^{-1} \left( \| f_Q \|_{\infty} \|Q - \hat{Q}\|_{L^2_{\mu}}\right)^{\frac{2}{3}}.
\end{equation}

\section{Pointwise norm bounds for partial evaluation}\label{apx:partial_eval}
\begin{lemma}\label{lemma:partial_eval}
    Let $f \in V \subseteq L^2_\mu\left([0, 1]^d\right)$, and suppose $\{P_\lambda\}_{\lambda \in \Lambda}$ is a tensorized orthonormal basis of $V$, with index set $\Lambda \subseteq \N_0^d$.
    Consider a subset $I \subseteq \{1, \ldots, d\}$ of indices, let $x \in (0, 1)^{|I|}$, and denote by $f_x$ the partial evaluation of $f$ in $x$ at indices given by $I$. Then, there are constants $C_1, C_2 > 0$, such that the $L^2$ norm of $f_x$ satisfies the inequality
    \begin{equation}\label{eq:partial_eval_norm_bound}
        \|f_x\|^2_{L^2_\mu\left([0, 1]^{d-|I|}\right)} < C_1 \left(C_2 + \|f\|^2_{L^2_\mu\left([0, 1]^d\right)}\right).
    \end{equation}
\end{lemma}
\begin{pf}
    Evaluating $f = \sum_{\lambda \in \Lambda} c_\lambda P_\lambda$ in $x$ at the coordinates corresponding to $I$ gives
    \begin{equation}
        f_x = \sum_{\lambda \in \Lambda} c_\lambda \prod_{i \in I} P_{\lambda_i}(x_i) \prod_{i \notin I} P_{\lambda_i},
    \end{equation}
    which has the reduced index set
    \begin{equation}
        \Lambda^I := \{(\lambda_i)_{i \notin I} : \lambda \in \Lambda\}.
    \end{equation}
    To compute the coefficients of this representation, one needs to collect all indices, that match on the complement of $I$ in $\{1, \ldots, d\}$.
    For instance, the coefficient $\hat{c}_\mu$ for index $\mu \in \Lambda^I$ is given by
    \begin{equation}
        \hat{c}_\mu = \sum_{\lambda \in \Lambda(\mu)} c_\lambda \prod_{i \in I} P_{\lambda_i}(x_i),
    \end{equation}
    where $\Lambda(\mu) = \{\lambda \in \Lambda : (\lambda_i)_{i \notin I} = \mu\}$.
    Note that the set of Legendre polynomials $\{P_n\}_{n \in \N_0}$ is orthonormal on the unit interval $[0, 1]$ with respect to the one-dimensional Lebesgue measure and therefore $\{P_\lambda\}_{\lambda \in \Lambda}$, where
    \begin{equation}
        P_\lambda(x) = \prod_{i=1}^d P_{\lambda_i}(x_i),
    \end{equation}
    forms an $L^2_\mu$ orthonormal basis of $V$.
    Moreover, Parseval's identity
    \begin{equation}
        \|f\|^2_{L^2_\mu\left([0, 1]^d\right)} = \sum_{\lambda \in \Lambda} c_\lambda^2
    \end{equation}
    holds.
    For the squared coefficients of the partial evaluation
    \begin{align}\label{eq:bound_coef}
        \begin{split}
            \hat{c}_\mu^2
            &= \left(\sum_{\lambda \in \Lambda(\mu)} c_\lambda \prod_{i \in I} P_{\lambda_i}(x_i)\right)^2 \\
            &= \sum_{\lambda, \nu \in \Lambda(\mu)} c_\lambda \, c_\nu \prod_{i \in I} P_{\lambda_i}(x_i) \, P_{\nu_i}(x_i) \\
            &< \underbrace{\left(\frac{2}{\pi}\right)^{|I|} \prod_{i \in I} \frac{1}{\sqrt{x_i (1 - x_i)}}}_{=: C_1} \, \sum_{\lambda, \nu \in \Lambda(\mu)} |c_\lambda||c_\nu|,
        \end{split}
    \end{align}
    where we used a sharpened Bernstein-type inequality for Legendre polynomials \cite{Antonov1981} in the last step,
    \begin{equation}
        |P_n(x)| < \left(\frac{2}{\pi} \frac{1}{\sqrt{x (1 - x)}}\right)^\frac{1}{2}, \quad x \in (0, 1), \quad \forall n \in \N_0.
    \end{equation}
    Applying Parseval's identity to the partial evaluation and using Ineq. \eqref{eq:bound_coef} yields
    \begin{equation}\label{eq:remark_helper}
        \|f_x\|^2_{L^2_\mu\left([0, 1]^{d-|I|}\right)} = \sum_{\mu \in \Lambda^I} \hat{c}_\mu^2 < C_1 \sum_{\mu \in \Lambda^I} \sum_{\lambda, \nu \in \Lambda(\mu)} |c_\lambda||c_\nu|.
    \end{equation}
    By splitting the inner sum into contributions from equal and unequal indices, and noting that $\bigcup_{\mu \in \Lambda^I} \Lambda(\mu) = \Lambda$, we have
    \begin{equation}
        \sum_{\mu \in \Lambda^I} \sum_{\lambda, \nu \in \Lambda(\mu)} |c_\lambda||c_\nu| = \underbrace{\sum_{\mu \in \Lambda^I} \sum_{\lambda \in \Lambda(\mu)} |c_\lambda|^2}_{ = \|f\|^2_{L^2_\mu\left([0, 1]^d\right)}} + \underbrace{\sum_{\mu \in \Lambda^I} \, \sum_{\substack{\lambda, \nu \in \Lambda(\mu) \\ \lambda \neq \nu}} |c_\lambda| |c_\nu|}_{=: C_2}.
    \end{equation}
    \qed
\end{pf}
\begin{remark}\label{rm:polynomial}
If the index set $\Lambda$ is finite, i.e., if $f$ is a polynomial, a more satisfying result can be achieved.
By applying the Cauchy–Schwarz inequality to the double sum in Ineq. \eqref{eq:remark_helper}, we obtain
\begin{equation}
    \|f_x\|^2_{L^2_\mu\left([0, 1]^{d-|I|}\right)} < C_1 \max_{\mu \in \Lambda^I} |\Lambda(\mu)| \, \cdot \|f\|^2_{L^2_\mu\left([0, 1]^d\right)},
\end{equation}
where $|\Lambda(\mu)|$ is finite since $\Lambda(\mu) \subset \Lambda$.
This result is advantageous because it removes the additive constant $C_2$, transferring full error control of, for example, $f = Q - \hat{Q}$, to partial evaluations in the geometry.
\end{remark}

\begin{figure}[h!]
    \centering
    \includegraphics[width=.5\linewidth]{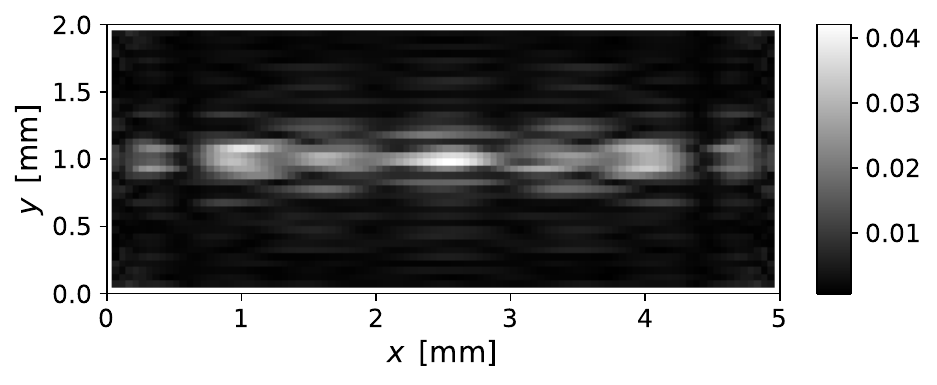}
    \caption{Visualization of the ratios $R(x, y)$, according to Eq. \eqref{eq:ratios}, across the considered channel, to in order to verify the bound in Ineq. \eqref{eq:partial_eval_norm_bound}.}
    \label{fig:ratios}
\end{figure}

In the remaining section, we aim to verify Ineq. \eqref{eq:partial_eval_norm_bound} from the previous Lemma \ref{lemma:partial_eval} for our specific setting, that is $f=Q-\hat{Q}$.
We utilize the surrogate $\hat{Q}$ from Section \ref{sec:full_channel_learning}.
The channel domain $[0, w] \times [0, h]$ is discretized using a uniform grid of $100 \times 40$ points.
For each grid point $(x, y)$, we compute the partial evaluation $\hat{Q}_{(x, y)}$, and the corresponding error $\|Q_{(x, y)} - \hat{Q}_{(x, y)}\|_{L^2_{\mu_\omega}}$ is approximated using a quasi-Monte Carlo method with $N=100$ samples.
Figure \ref{fig:ratios} down below, displays
\begin{equation}\label{eq:ratios}
    R(x, y) = \left(\frac{\|Q_{(x, y)} - \hat{Q}_{(x, y)}\|^2_{L^2_{\mu_\omega}}}{C_1 \left(C_2 + \|Q - \hat{Q}\|^2_{L^2_\mu}\right)}\right)^{1/2}
\end{equation}
across the $100 \times 40$ grid.
Since all ratios are smaller than $1$, we conclude that Ineq. \eqref{eq:partial_eval_norm_bound} applies to all of the $4000$ partial evaluations in the channel considered here.
Note that $C_2$ depends on the Legendre coefficients of the function at hand.
The higher the regularity of the function, the more rapidly the coefficients decay, ensuring that $C_2$ remains relatively small.
Here, $C_2 \approx 0.023$, where the Legendre coefficients of $Q-\hat{Q}$ are approximated using quasi-Monte Carlo integration with $N=2 \cdot 10^6$ samples.

\section{Complexity analysis}\label{apx:complexity}
\begin{figure*}[h!]
  \centering
   \includegraphics[width=\textwidth]{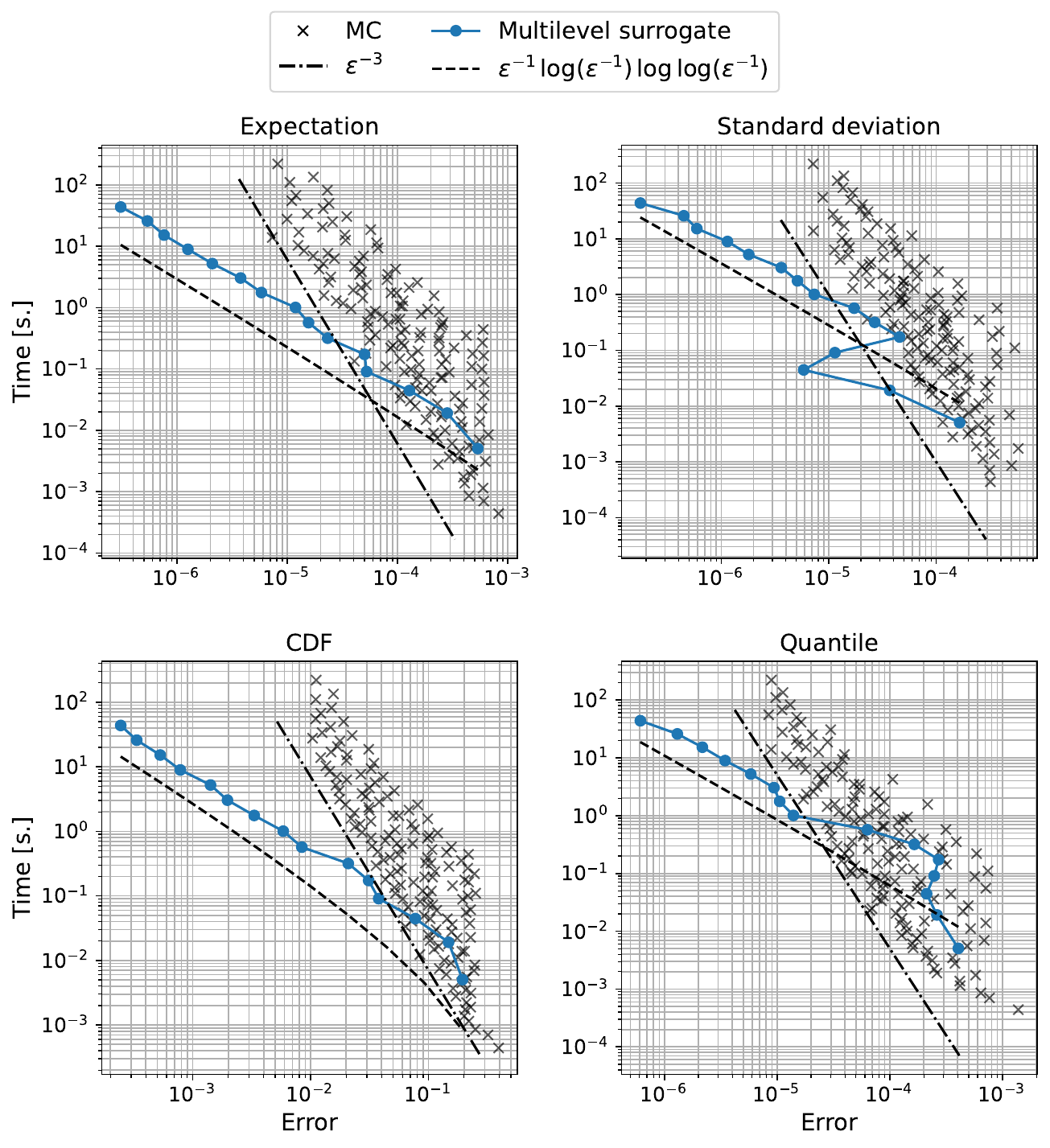}
    \caption{Estimated errors of the corresponding statistical quantities for the fractional anisotropy vs. runtime. Each colored dot corresponds to a multilevel polynomial surrogate $\hat{Q}_\epsilon$ with a specific tolerance $\epsilon > 0$.
    The crosses correspond to standard Monte Carlo estimations. All computations were performed using the FTE, Eq. \eqref{eq:fte}.}
    \label{fig:complexity_plot}
\vspace{1cm}
\end{figure*}

\begin{figure*}[h!]
  \centering
   \includegraphics[width=\textwidth]{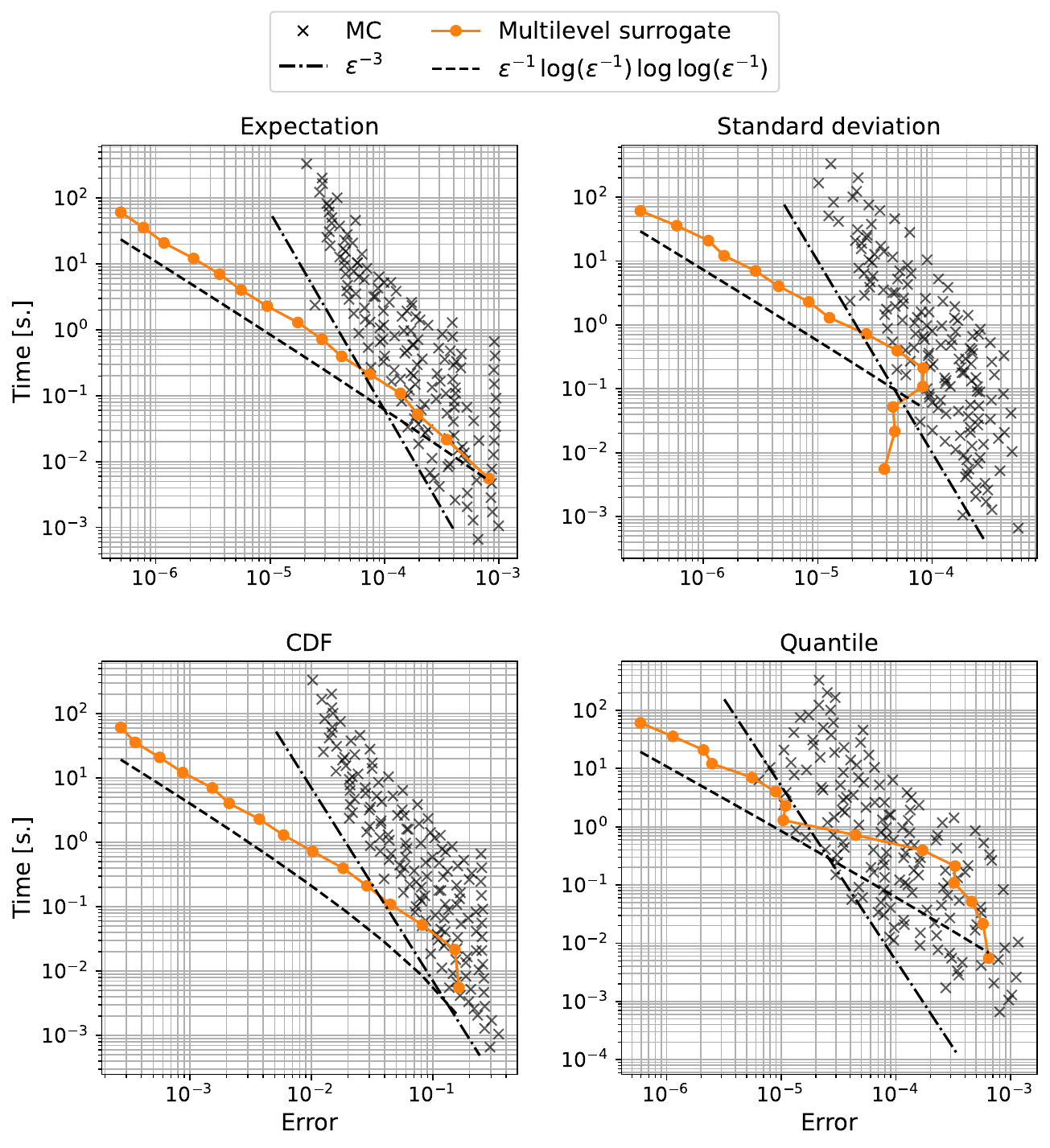}
    \caption{Estimated errors of the corresponding statistical quantities for the fractional anisotropy vs. runtime. Each colored dot corresponds to a multilevel polynomial surrogate $\hat{Q}_\epsilon$ with a specific tolerance $\epsilon > 0$.
    The crosses correspond to standard Monte Carlo estimations. All computations were performed using the iARD model, Eq. \eqref{eq:ard}.}
    \label{fig:complexity_plot_iard}
\end{figure*}

In the following, we outline the asymptotic computational costs of the multilevel surrogate method and compare to those of standard Monte Carlo.
Utilizing the forward Euler method and tensor product polynomial spaces corresponds to the rates $\beta = 1$, $\gamma = 1$, $\alpha = 3$, and $\sigma = 2$ in \cite{HajiAli2020}.
With these rates, the computational costs for constructing $\hat{Q}_\epsilon$ given some error tolerance $\epsilon$ behave asymptotically like
\begin{equation}
    \epsilon^{-1} \log (\epsilon^{-1}) \log \log (\epsilon^{-1}).
\end{equation}
This can be observed in Figure \ref{fig:complexity_plot}, where the runtime for computing statistical quantities of the QoI is compared to the resulting error.
All experiments were conducted using an Intel Core i5-1335U.
It can be observed, that the solid blue curve, corresponding to the multilevel polynomial surrogate approach, is parallel to the dashed curve with the predicted computational costs.
The surrogate-based approach is compared to the standard Monte Carlo method.
Each cross in the same figure corresponds to one Monte Carlo estimation using $N \in \N$ samples and a number of time steps $n \in \N$ for the forward Euler method.
In particular, ten logarithmically distributed sample sizes $N$ between $10^1$ and $10^4$, as well as fifteen logarithmically distributed discretization points $n \in \mathbb{N}$ between $10^{1.3}$ and $10^4$ were used.
One can observe the typical Monte Carlo rate of $\mathcal{O}(\epsilon^{-3})$ and the efficiency increase when using the multilevel polynomial surrogate method.
Figure \ref{fig:complexity_plot_iard} shows the same plot with the iARD model used as FOM.

\newpage
\section{Parameters for the numerical experiments}\label{apx:params}
\begin{table}[pos=h!]
\caption{Cross-WLF parameter values used for the numerical experiments corresponding to a 30 wt\% glass fiber filled material \cite{Moldflow2023}.}\label{tbl:basf}
\begin{tabular*}{\tblwidth}{@{}LLL@{}}
\toprule
Parameter & Value & Unit\\
\midrule
$n$ & $0.3267$ & - \\
$\dot{\gamma}$& 1 & 1 / s \\
$\tau^*$ & $123991$ & Pa\\
$T^*$ & $323.15$ & K\\
$D$ & $2.06635 \cdot 10^{15}$ & Pa $\cdot$ s\\
$A_1$ & $36.07$ & -\\
$A_2$ & $51.6$ & K\\
\bottomrule
\end{tabular*}
\end{table}

\begin{table}[pos=h!]
\caption{Further parameters for the numerical experiments.}\label{tbl:params}
\begin{tabular*}{\tblwidth}{@{}LLLL@{}}
\toprule
Parameter & Description & Value & Unit\\ 
\midrule
$\Omega_T$ & Temperature domain  & $[550, 580]$ & K\\
$\Omega_{L_f}$ & Fiber length domain & $[0.38 \pm 50\%]$ & mm\\
$(w, h)$ & Channel width and height & $[5, 2]$ & mm\\
$(x_0, y_0)$ & Arbitrary selected point & $[2.0, 0.1]$ &-\\
$P$ & Pressure gradient & 10 & Pa\\
$C_I$ & Fiber interaction coefficient & 0.01 & -\\
$C_M$ & Anisotropy coefficient & 0.2 & -\\
$D_f$ & Fiber diameter & 0.015 & mm\\
$t_\text{end}$ & Time horizon for FOM & 300 & s\\
$p$ & Quantile probability & $0.99$ & -\\
$f$ & Hybrid closure parameter & 1 & - \\
\bottomrule
\end{tabular*}
\end{table}

\twocolumn[
\begin{center}
\addcontentsline{toc}{section}{References}
\end{center}
]

\section*{Acknowledgments}
We would like to thank MathSEE for the Bridge PhD funding and gratefully acknowledge the support of the Deutsche Forschungsgemeinschaft (DFG) for funding the research project "MeproSi" (project number: 464119659), which has enabled and informed the scope of this work.

\printcredits

\bibliography{bibliography}

\end{document}